\documentclass[longauth]{aa}  

\usepackage{graphicx}
\usepackage{color}
\usepackage{epstopdf}
\usepackage{txfonts}
\usepackage{natbib}
\bibpunct{(}{)}{;}{a}{}{,} 

\newcommand{\g}{$\gamma$}
\newcommand{\wco}{W_\mathrm{CO}}
\newcommand{\hi}{\ion{H}{i}}
\newcommand{\hii}{\ion{H}{ii}}
\newcommand{\hd}{\mathrm{H}_2}
\newcommand{\nhi}{N(\hi)}
\newcommand{\nhii}{N(\hii)}
\newcommand{\ebv}{\mathrm{E(B-V)}}
\newcommand{\av}{\mathrm{A_V}}
\newcommand{\qhi}[1]{q_{\mathrm{H\,\scriptscriptstyle{I}} , \, #1}}
\newcommand{\qco}[1]{q_{\mathrm{CO} , \, #1}}
\newcommand{\qdust}{q_\mathrm{\av}}
\newcommand{\xco}{X_\mathrm{CO}}
\newcommand{\xav}{X_\mathrm{A_V}}
\newcommand{\avres}{\mathrm{A}_\mathrm{V,exc}}
\newcommand{\nhd}{N(\mathrm{H_2})}

\newcommand{\qhinull}{q_\mathrm{H\,\scriptscriptstyle{I}}}

\newcommand{\astropp}{Astroparticle Physics}

\begin{document}



\titlerunning{Interstellar \g-ray emission from Cygnus with \textit{Fermi}}

\authorrunning{The \textit{Fermi} LAT collaboration}

\pagestyle{empty}

\title{The cosmic-ray and gas content of the Cygnus region as measured in gamma rays by
the \textit{Fermi} Large Area Telescope
}
\author{
M.~Ackermann$^{(1)}$ \and 
M.~Ajello$^{(1)}$ \and 
A.~Allafort$^{(1)}$ \and 
L.~Baldini$^{(2)}$ \and 
J.~Ballet$^{(3)}$ \and 
G.~Barbiellini$^{(4,5)}$ \and 
D.~Bastieri$^{(6,7)}$ \and 
A.~Belfiore$^{(8)}$ \and 
R.~Bellazzini$^{(2)}$ \and 
B.~Berenji$^{(1)}$
% ...
}
\authorrunning{LAT collaboration}

\institute{
\inst{1}~W. W. Hansen Experimental Physics Laboratory, Kavli Institute for Particle Astrophysics and
Cosmology, Department of Physics and SLAC National Accelerator Laboratory, Stanford University,
Stanford, CA 94305, USA\\ 
\inst{2}~Istituto Nazionale di Fisica Nucleare, Sezione di Pisa, I-56127 Pisa, Italy\\ 
\inst{3}~Laboratoire AIM, CEA-IRFU/CNRS/Universit\'e Paris Diderot, Service d'Astrophysique, CEA
Saclay, 91191 Gif sur Yvette, France\\ 
\inst{4}~Istituto Nazionale di Fisica Nucleare, Sezione di Trieste, I-34127 Trieste, Italy\\ 
\inst{5}~Dipartimento di Fisica, Universit\`a di Trieste, I-34127 Trieste, Italy\\ 
\inst{6}~Istituto Nazionale di Fisica Nucleare, Sezione di Padova, I-35131 Padova, Italy\\ 
\inst{7}~Dipartimento di Fisica ``G. Galilei", Universit\`a di Padova, I-35131 Padova, Italy\\ 
\inst{8}~INAF-Istituto di Astrofisica Spaziale e Fisica Cosmica, I-20133 Milano, Italy\\ 
\inst{9}~Istituto Nazionale di Fisica Nucleare, Sezione di Perugia, I-06123 Perugia, Italy\\ 
\inst{10}~Dipartimento di Fisica, Universit\`a degli Studi di Perugia, I-06123 Perugia, Italy\\ 
\inst{11}~Dipartimento di Fisica ``M. Merlin" dell'Universit\`a e del Politecnico di Bari, I-70126
Bari, Italy\\ 
\inst{12}~Istituto Nazionale di Fisica Nucleare, Sezione di Bari, 70126 Bari, Italy\\ 
\inst{13}~Laboratoire Leprince-Ringuet, \'Ecole polytechnique, CNRS/IN2P3, Palaiseau, France\\ 
\inst{14}~Institut de Ci\`encies de l'Espai (IEEE-CSIC), Campus UAB, 08193 Barcelona, Spain\\ 
\inst{15}~Artep Inc., 2922 Excelsior Springs Court, Ellicott City, MD 21042, resident at Naval
Research Laboratory, Washington, DC 20375\\ 
\inst{16}~ASI Science Data Center, I-00044 Frascati (Roma), Italy\\ 
\inst{17}~Laboratoire Univers et Particules de Montpellier, Universit\'e Montpellier 2, CNRS/IN2P3,
Montpellier, France\\ 
\inst{18}~Dipartimento di Fisica, Universit\`a di Udine and Istituto Nazionale di Fisica Nucleare,
Sezione di Trieste, Gruppo Collegato di Udine, I-33100 Udine, Italy\\ 
\inst{19}~Space Science Division, Naval Research Laboratory, Washington, DC 20375-5352\\ 
\inst{20}~Universit\'e Bordeaux 1, CNRS/IN2p3, Centre d'\'Etudes Nucl\'eaires de Bordeaux Gradignan,
33175 Gradignan, France\\ 
\inst{21}~Department of Physical Sciences, Hiroshima University, Higashi-Hiroshima, Hiroshima
739-8526, Japan\\ 
\inst{22}~INAF Istituto di Radioastronomia, 40129 Bologna, Italy\\ 
\inst{23}~Current address: Max-Planck-Institut f\"ur Radioastronomie, Auf dem H\"ugel 69, 53121
Bonn, Germany\\ 
\inst{24}~Center for Space Plasma and Aeronomic Research (CSPAR), University of Alabama in
Huntsville, Huntsville, AL 35899\\ 
\inst{25}~NASA Goddard Space Flight Center, Greenbelt, MD 20771, USA\\ 
\inst{26}~Science Institute, University of Iceland, IS-107 Reykjavik, Iceland\\ 
\inst{27}~College of Science , Ibaraki University, 2-1-1, Bunkyo, Mito 310-8512, Japan\\ 
\inst{28}~Research Institute for Science and Engineering, Waseda University, 3-4-1, Okubo, Shinjuku,
Tokyo 169-8555, Japan\\ 
\inst{29}~CNRS, IRAP, F-31028 Toulouse cedex 4, France\\ 
\inst{30}~Universit\'e de Toulouse, UPS-OMP, IRAP, Toulouse, France\\ 
\inst{31}~Yukawa Institute for Theoretical Physics, Kyoto University, Kitashirakawa Oiwake-cho,
Sakyo-ku, Kyoto 606-8502, Japan\\ 
\inst{32}~Max-Planck Institut f\"ur extraterrestrische Physik, 85748 Garching, Germany\\ 
\inst{33}~Department of Physics and Department of Astronomy, University of Maryland, College Park,
MD 20742\\ 
\inst{34}~Istituto Nazionale di Fisica Nucleare, Sezione di Roma ``Tor Vergata", I-00133 Roma,
Italy\\ 
\inst{35}~Department of Physics, Boise State University, Boise, ID 83725, USA\\ 
\inst{36}~Hiroshima Astrophysical Science Center, Hiroshima University, Higashi-Hiroshima, Hiroshima
739-8526, Japan\\ 
\inst{37}~Institute of Space and Astronautical Science, JAXA, 3-1-1 Yoshinodai, Chuo-ku, Sagamihara,
Kanagawa 252-5210, Japan\\ 
\inst{38}~Department of Physics and Astronomy, University of Denver, Denver, CO 80208, USA\\ 
\inst{39}~Max-Planck-Institut f\"ur Physik, D-80805 M\"unchen, Germany\\ 
\inst{40}~Center for Earth Observing and Space Research, College of Science, George Mason
University, Fairfax, VA 22030, resident at Naval Research Laboratory, Washington, DC 20375\\ 
\inst{41}~Institut f\"ur Astro- und Teilchenphysik and Institut f\"ur Theoretische Physik,
Leopold-Franzens-Universit\"at Innsbruck, A-6020 Innsbruck, Austria\\ 
\inst{42}~Santa Cruz Institute for Particle Physics, Department of Physics and Department of
Astronomy and Astrophysics, University of California at Santa Cruz, Santa Cruz, CA 95064, USA\\ 
\inst{43}~NYCB Real-Time Computing Inc., Lattingtown, NY 11560-1025, USA\\ 
\inst{44}~Department of Physics, Center for Cosmology and Astro-Particle Physics, The Ohio State
University, Columbus, OH 43210, USA\\ 
\inst{45}~Partially supported by the International Doctorate on Astroparticle Physics (IDAPP)
program\\ 
\inst{46}~Instituci\'o Catalana de Recerca i Estudis Avan\c{c}ats (ICREA), Barcelona, Spain\\ 
\inst{47}~Consorzio Interuniversitario per la Fisica Spaziale (CIFS), I-10133 Torino, Italy\\ 
\inst{48}~INTEGRAL Science Data Centre, CH-1290 Versoix, Switzerland\\ 
\inst{49}~NASA Postdoctoral Program Fellow, USA\\ 
\inst{50}~Dipartimento di Fisica, Universit\`a di Roma ``Tor Vergata", I-00133 Roma, Italy\\ 
\inst{51}~Department of Physics, Stockholm University, AlbaNova, SE-106 91 Stockholm, Sweden\\ 
\inst{52}~The Oskar Klein Centre for Cosmoparticle Physics, AlbaNova, SE-106 91 Stockholm, Sweden\\ 
\inst{53}~Laboratoire d'Astrophysique de Bordeaux, Universit\'e de Bordeaux, CNRS/INSU, Floirac
cedex, France\\ 
\email{isabelle.grenier@cea.fr} \\
\email{luigi.tibaldo@pd.infn.it} \\
}

\date{Received \ldots; accepted \ldots}

\abstract
{{ The Cygnus region hosts a giant molecular-cloud complex that actively forms massive
stars}. Interactions of
cosmic rays with interstellar gas and radiation fields make it shine at \g-ray energies. Several
\g-ray pulsars and other energetic
sources are seen in this direction.}
{ { In this paper we analyze the \g-ray emission
measured by the \textit{Fermi} Large Area Telescope in the energy range from
100~MeV to 100~GeV in order to probe the gas and cosmic-ray content on the scale of
the whole Cygnus complex. The \g-ray emission on the scale of the
central massive stellar clusters and from individual sources is
addressed elsewhere.}}
{The signal from bright pulsars is greatly reduced by selecting photons in their off-pulse phase
intervals. We compare the diffuse \g-ray emission with interstellar gas maps derived
from radio/mm-wave lines and visual extinction data. A general model of the region, including
other pulsars and \g-ray sources, is sought.}
{The integral \hi\ emissivity above $100$~MeV {{  averaged over the whole Cygnus complex}}
amounts to $[2.06
\pm 0.11 \,(\mathrm{stat.})\; ^{+0.15}
_{-0.84}\,(\mathrm{syst.})]\times 10^{-26}$ photons s$^{-1}$ sr$^{-1}$ H-atom$^{-1}$, where the
systematic error is dominated by the uncertainty on the $\hi$ opacity to calculate its column
densities. {{  The integral
emissivity and its
spectral
energy distribution}} are both consistent within the systematics with LAT measurements in the
interstellar
space near the solar system.
The {{  average}} $\xco=\nhd/\wco$ ratio is found to be
$[1.68 \pm 0.05\,(\mathrm{stat.})\; _{-0.10}^{+0.87}
\,(\hi\;\mathrm{opacity})] \times 10^{20}$
molecules cm$^{-2}$ (K km s$^{-1}$)$^{-1}$, consistent with other LAT measurements in the
Local Arm. We
detect significant \g-ray emission from dark neutral gas for a mass corresponding to $\sim 40\%$ of
what is
traced by CO. The total interstellar mass in the Cygnus complex inferred from its \g-ray emission
amounts to
$8\,^{+5}_{-1}\times 10^{6} M\sun$ at a distance of 1.4 kpc.}
{Despite the conspicuous star formation activity and high masses of the interstellar clouds, the
cosmic-ray population in the Cygnus complex averaged over a few hundred parsecs is similar to that
of
the local interstellar space.}


\maketitle


\section{Introduction}
Regions with conspicuous star formation activity are of great interest for
understanding the life cycle of interstellar matter and the properties of cosmic rays (CRs) in the
Galaxy. Interstellar \g-ray emission produced by CR interactions with the interstellar gas via
nucleon-nucleon inelastic collisions and electron Bremsstrahlung can be used to probe their CR
and gas content.

High-energy \g-ray observations have entered a new era since the launch of the \textit{Fermi
Gamma-ray Space Telescope} in 2008. The \textit{Fermi} Large Area Telescope
\citep[LAT;][]{atwood2009} has already measured strong \g-ray emission 
toward the 30 Doradus starburst region in the Large Magellanic Cloud
\citep{abdo2010lmc}, { and it also} pointed out a global correlation between the \g-ray
luminosity and
star-formation rate
in a few normal galaxies \citep{abdo2010starbust}.

A primary observational target for \textit{Fermi} in our Galaxy is the Cygnus~X star-forming region,
owing to its
proximity \citep[$\sim 1.4$~kpc;][]{hanson2003,negueruela2008} and the
availability of numerous multiwavelength observations. Named
after the strong emission at
X-ray
wavelengths \citep{cash1980}, Cygnus~X is
located
around the Galactic longitude $l=80\degr$, tangent to the Local Spur. It contains
numerous \ion{H}{ii}
regions and OB
associations \citep{uyaniker2001,leduigou2002}. It has long been debated whether it represents a
coherent complex
or the alignment of different structures along the line of
sight. Recent high-resolution observations by \citet{schneider2006} and \citet{roy2011} have pointed
out that most of the molecular clouds in the Cygnus~X region are connected and partly
show evidence
of interactions with the massive stellar cluster Cygnus OB2 and other OB associations in the
region. Foreground molecular clouds from the Great Cygnus Rift, at 0.6--0.8 kpc, contribute little
to the high mass seen { in interaction with} the Cygnus X region itself, at 1.4 kpc.
Therefore, the
molecular cloud complex
appears as one of the most massive in
the Galaxy. Atomic gas seen in these directions is probably more widespread along the line of
sight.

\citet{abdo2007,abdo2008milagrodiff} analyzed Milagro measurements at energies $>10$~TeV
and reports an excess of diffuse \g-ray emission with respect to predictions based on CR spectra
equivalent to those near the Earth. They attribute the excess to the
possible presence of freshly accelerated particles. 

{ The escape of CRs from their sources and the early
propagation in the surrounding medium have so far been poorly constrained by observations. In
particular,
particles
accelerated in regions of massive star-formation are likely to be significantly influenced by
the turbulent environment. It is therefore interesting to investigate how the CR populations on the
scale of the massive stellar clusters and on the larger scale of the parent interstellar complex
compare with each other and with the average CR population of the Local Spur.}

{ This paper reports our analysis of the \g-ray emission measured
by the LAT
in the energy range
between 100~MeV and 100~GeV across the entire Cygnus
region}. {  We focus on the  large-scale properties of the
interstellar
emission to probe
the CR population and to complement gas and dust
observations at other wavelengths to constrain the amount of gas in different phases
over the whole Cygnus complex}. We also build an
improved interstellar background framework for the study of individual \g-ray sources that will be
treated in
companion papers. { We discuss interstellar emission in the star-forming
region of Cygnus X in a dedicated
paper \citep{cocoonScience}.}

\section{Data}
{
In this section we  describe the data used in the paper. First of all, we  describe
the \g-ray data sample (\S~\ref{gammadata}), including the selection criteria (\S~\ref{datasel}) and
the
procedure to mitigate the signals from the brightest pulsars (\S~\ref{psr}). Then, we  present
an overview of the other data used to trace the distribution of the interstellar matter
(\S~\ref{otherdata}), including
radio and mm-wave lines (\S~\ref{linedata}) to trace neutral gas, visual extinction
(\S~\ref{avmaps}) to
trace dark neutral gas, and the free-free emission intensities obtained from microwave observations
(\S~\ref{freefreemap}) to trace the ionized gas.
}

\subsection{Gamma-ray data}\label{gammadata}

\subsubsection{Observations and data selection}\label{datasel}

The LAT is a pair-tracking telescope detecting photons from 20~MeV to more than 300~GeV. The
instrument is described in \citet{atwood2009} and its on-orbit calibration in
\citet{abdo2009orbcal}. The LAT operates most of the time in continuous sky-survey mode. We
accumulated data for our
region of interest from August 5, 2008 (MET\footnote{\textit{Fermi} Mission Elapsed Time, i.e.
seconds since 2001 January 1 at 00:00:00 UTC.}  239587201)
to August 5, { 2010} (MET 302659202).

We selected data according to the tightest available background rejection criteria,
corresponding to the
\emph{Pass~6 Dataclean} event class \citep{abdo2010isoback}\footnote{{ Performance figures
for the \emph{Dataclean} event selection are given in the
reference.}}. In order to limit the contamination from
the
Earth's atmospheric \g-ray
emission, we selected events with measured arrival
directions within $100\degr$ of the local zenith and within $65\degr$ of the instrument
boresight, taken during periods when the LAT rocking
angle was less than $52 \degr$.

The angular resolution of the LAT strongly depends on the photon energy, improving as the
energy increases \citep{atwood2009}. Confusion at low energies is a problem since we aim to
spatially separate the different components in the crowded Cygnus~X region. We therefore accepted
below 1
GeV
only photons
that produced electron-positron pairs in the thin
converter planes of the \emph{front} section of the tracker, which provides a higher angular
resolution \citep{atwood2009}. Above 1 GeV, we kept all events which converted either in the 
\emph{front} or \emph{back} section of the tracker.

We analysed data at Galactic longitudes $72\degr \leq l \leq 88\degr$ and latitudes
$-15\degr
\leq b \leq +15\degr$. The longitude window contains the interstellar
complexes associated with
Cygnus~X; the latitude window is large enough to allow a reliable separation of the
large-scale emission from atomic gas, isotropic background and inverse-Compton (IC)
scattering of low-energy radiation fields by CR electrons. We analysed
the data in the
100~MeV--100~GeV energy band. Below 100~MeV the instrumental systematics are large
\citep{rando2009} and the angular resolution is poor, whereas above 100~GeV we are limited
by the low photon statistics.

\subsubsection{Removal of bright pulsars}\label{psr}
Three bright pulsars dominate the \g-ray emission from the region
below a few GeV: the radio pulsar J2021$+$3651 \citep{abdo20092021p3651}
and the two LAT-discovered pulsars J2021$+$4026 and J2032$+$4127 \citep{abdo2009blindpsr}.
To increase the sensitivity to faint sources and to the spatial structure of the diffuse emission,
we reduced their contribution
by excluding the periodic time intervals when their pulsed emission is the most intense.
{
Removing the intense pulsed flux helps to reduce the impact of any incorrect modeling of such
bright sources on the results.
} 

\begin{figure}
\resizebox{\hsize}{!}{\includegraphics{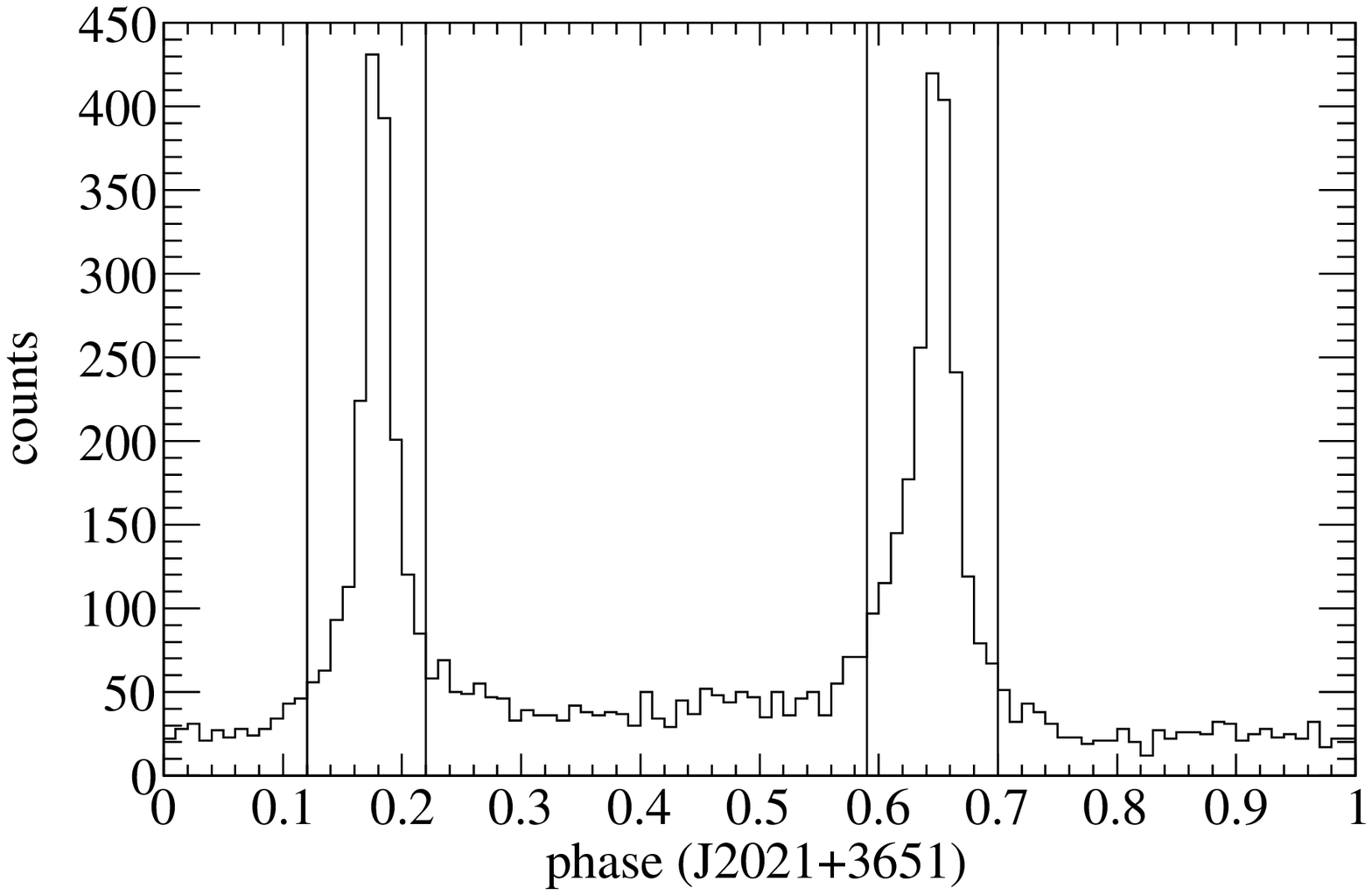}}
\resizebox{\hsize}{!}{\includegraphics{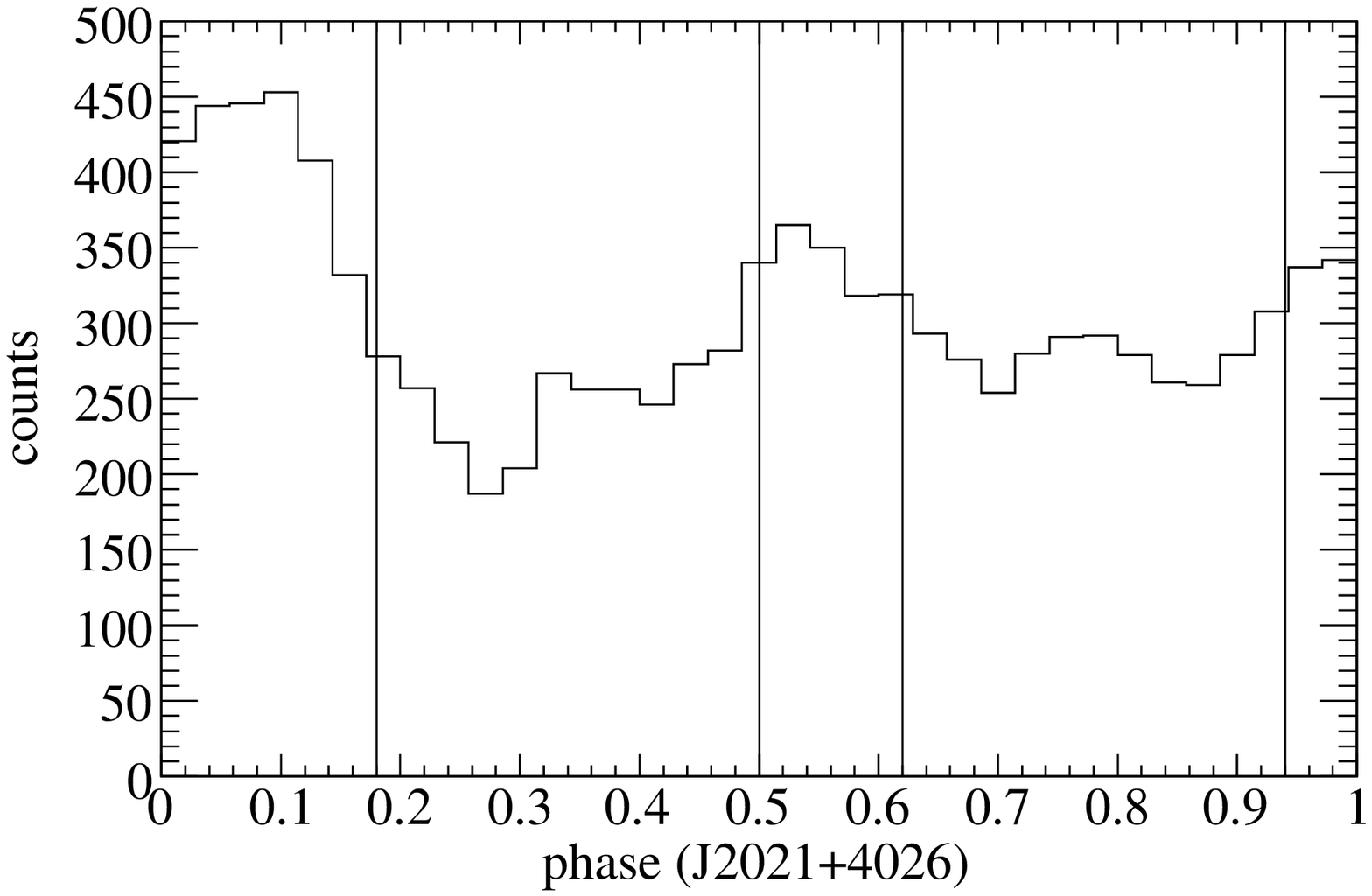}}
\resizebox{\hsize}{!}{\includegraphics{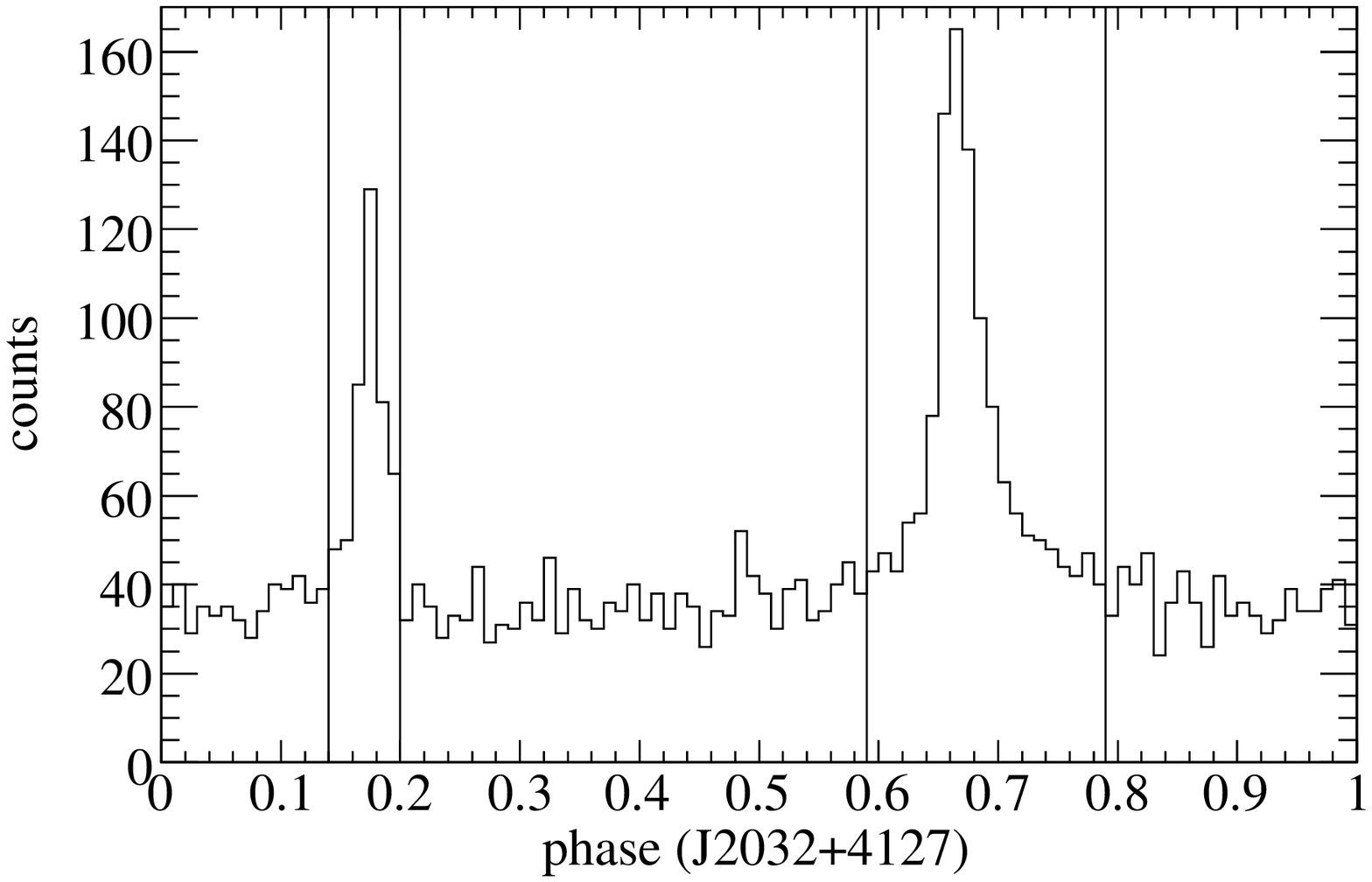}}
\caption{Light curves and off-pulse phase intervals for the three bright pulsars. The light curves
are
constructed for illustration purposes with photons recorded in a circular
region of radius $0.5 \degr$ around the pulsar positions and energies $>200$~MeV.}
\label{PSRlc}
\end{figure}

To assign pulse phases for each of the three pulsars, we produced
timing models using \textsc{Tempo2} \citep{hobbs2006} according to the method
described in \citet{ray2011}\footnote{For the three pulsars, the RMS of the timing residuals is
below
1.1\% of their rotational period.}. Figure~\ref{PSRlc}
shows the three light curves and the phase intervals with bright pulsed
emission. The phase boundaries are reported in Table~\ref{table:PSR}, together with the fraction of
time in the off-pulse interval suitable for our study.
There is a considerable level of off-pulse emission toward PSR~J2021$+$4026 that cannot be removed
\citep{abdo2010psrcat};
however, given the brightness of
the source, the removal of the on-pulse interval is useful for our aims.

\begin{table}
\caption{Off-pulse phase intervals and time fractions} of the three
bright
pulsars. 
\label{table:PSR} 
\centering 
\begin{tabular}{c c c} 
\hline\hline 
PSR & phase interval 	& time fraction (\%)\\ 
\hline 
J2021$+$3651 & 0--0.12, 0.22--0.59, 0.7--1 & 79\\
J2021$+$4026 & 0.18--0.5, 0.62--0.94 &  64\\
J2032$+$4127 & 0--0.14, 0.2--0.59, 0.79--1 & 74\\   
\hline 
\end{tabular}
\end{table}

\begin{figure}
\resizebox{\hsize}{!}{\includegraphics{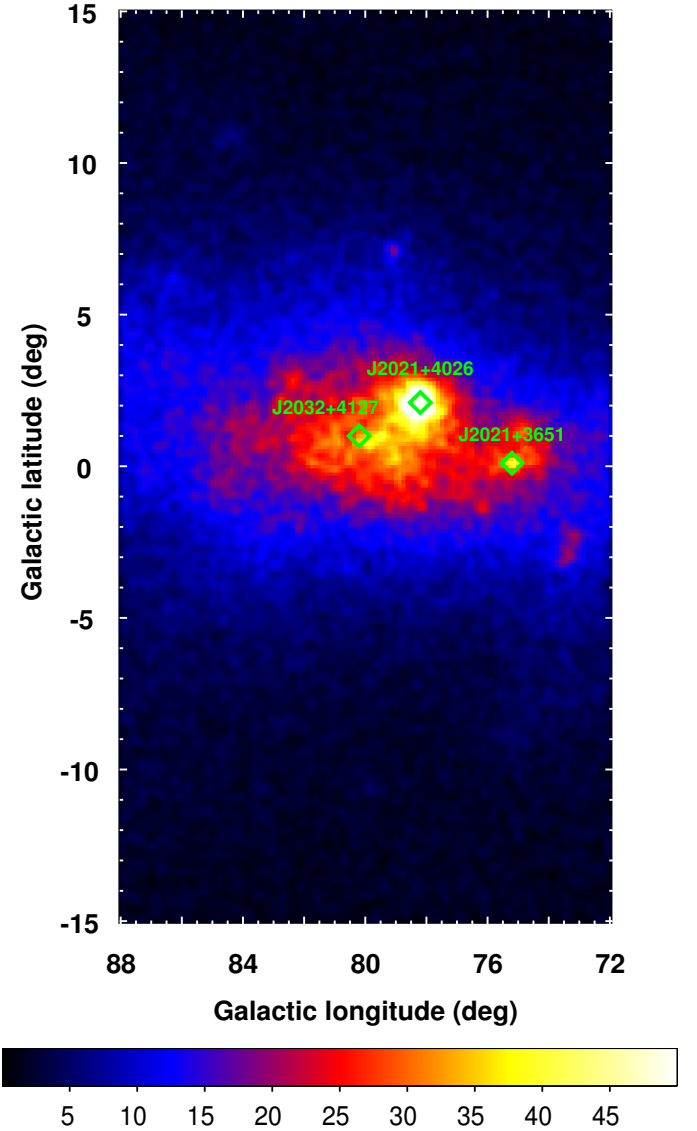}}
\caption{Total count map in the energy range 100 MeV--100 GeV, binned over a $0.125\degr \times
0.125\degr$ grid in Galactic coordinates in Cartesian projection. Data were selected according to
the criteria described in the text (\S~\ref{datasel}) and in the off-pulse phase
intervals
of the three bright pulsars (\S~\ref{psr}), whose positions are marked by diamonds. Counts
are saturated between 0 and { 50}, and smoothed for
display
with a Gaussian kernel of $\sigma=0.25\degr$.}
\label{countmap}
\end{figure}

A total count map in the off-pulse phase intervals of the three bright pulsars is provided for
illustration in Fig.~\ref{countmap}.
To remove the pulsar signal without excessively sacrificing the
photon statistics in other
directions, we restricted the timing selection to a
circular region around the pulsar position, namely to pixels in our angular grid (described
later in \S~\ref{anamethod})
the centroids of which lie within the energy-dependent radius 
\begin{equation}
 r_{\mathrm{cut}}(E)= 2 \cdot \left[0.8\degr \left(\frac{E}{1\:\mathrm{GeV}}\right)^{-0.8} \oplus
0.07\degr
\right]
\end{equation}
where the symbol $\oplus$ indicates addition in quadrature. This is an
approximate representation of the LAT $95\%$ containment angle as a function of energy.
We note that the accurate parametrization of the LAT point spread function (PSF) depends on
energy, pair-conversion point in the tracker and, to a lesser extent, on the incidence angle. The
PSF is best represented by the LAT instrument response functions (IRFs), which are used later for
the
likelihood analysis. The above
acceptance-averaged approximation for the containment angle is only useful
for calculating the radius $r_\mathrm{cut}$, and we verified that the results are insensitive to
reasonable
variations in this parameter.

To take the cut on pulsar phases into account, for each direction in the sky and energy the
exposure (see again \S~\ref{anamethod})
was multiplied by the remaining livetime fraction. The remainder of the pulsar emission was
included in the model using
\begin{itemize}
 \item a point source to represent emission in the off-pulse interval;
 \item a second point source, for which the number of expected counts is set to zero at $r <
r_{cut}(E)$ from the pulsar position, to represent on-pulse \g-rays spilling over at $r >
r_{cut}(E)$ in the tails of the PSF. 
\end{itemize}
The two sources have free independent fluxes in each energy bin of the analysis. This is
particularly important to account for the different spectra of the on-pulse and off-pulse
\g-ray emission and also to compensate for any mismatch between the
tails of the model PSF and the emission at large angles from the brightest sources
in the region.

Since the three pulsars have exponential spectral cutoffs near $2-3$~GeV
\citep{abdo2010psrcat} the { phase selection} was not applied above $10$~GeV 
where the level of pulsed emission is low { and each pulsar was accounted for by a single
point source}.  On the other hand, given the { abundant} statistics
but poor angular resolution at low energies (more than half of the region of interest would
be subject to on-pulse event
removal), we selected off-pulse photons for the whole region below $316$~MeV \footnote{See
\S~\ref{anamethod} for the definition of the energy grid used in the analysis.}. { In this
case no ``spill-over'' source was necessary.}

\subsection{Ancillary data}\label{otherdata}

\subsubsection{Radio/mm-wave lines: neutral gas}\label{linedata}

Neutral atomic hydrogen, \hi, was traced thanks to its 21-cm line. Where available\footnote{The CGPS
coverage is almost
complete at $-3.5\degr\leq b\leq +5.5\degr$ for this longitude range.} we used data from the
Canadian Galactic
Plane Survey \citep[CGPS;][]{taylor2003} rebinned onto the $0.125\degr \times
0.125\degr$ grid used for the other maps. Elsewhere, we used data from the
Leiden/Argentine/Bonn \citep[LAB;][]{kalberla2005} survey, with a coarser binning of $0.5\degr$. We
checked the consistency of the two survey calibrations in the overlap region.

Molecular hydrogen cannot be observed directly in its most abundant cold phase.
The velocity-integrated brightness temperature of the $^{12}$CO 2.6-mm line, $\wco$, is often
assumed to linearly scale with the $\nhd$ column density. We used CO data from the
composite survey by
\citet{dame2001}, filtered with the moment-masking technique \citep{dame2011} in order to reduce the
noise while                                                              
preserving the faint cloud edges and 
keeping the resolution of the original data.

\begin{figure*}
\centering
\includegraphics[width=17cm]{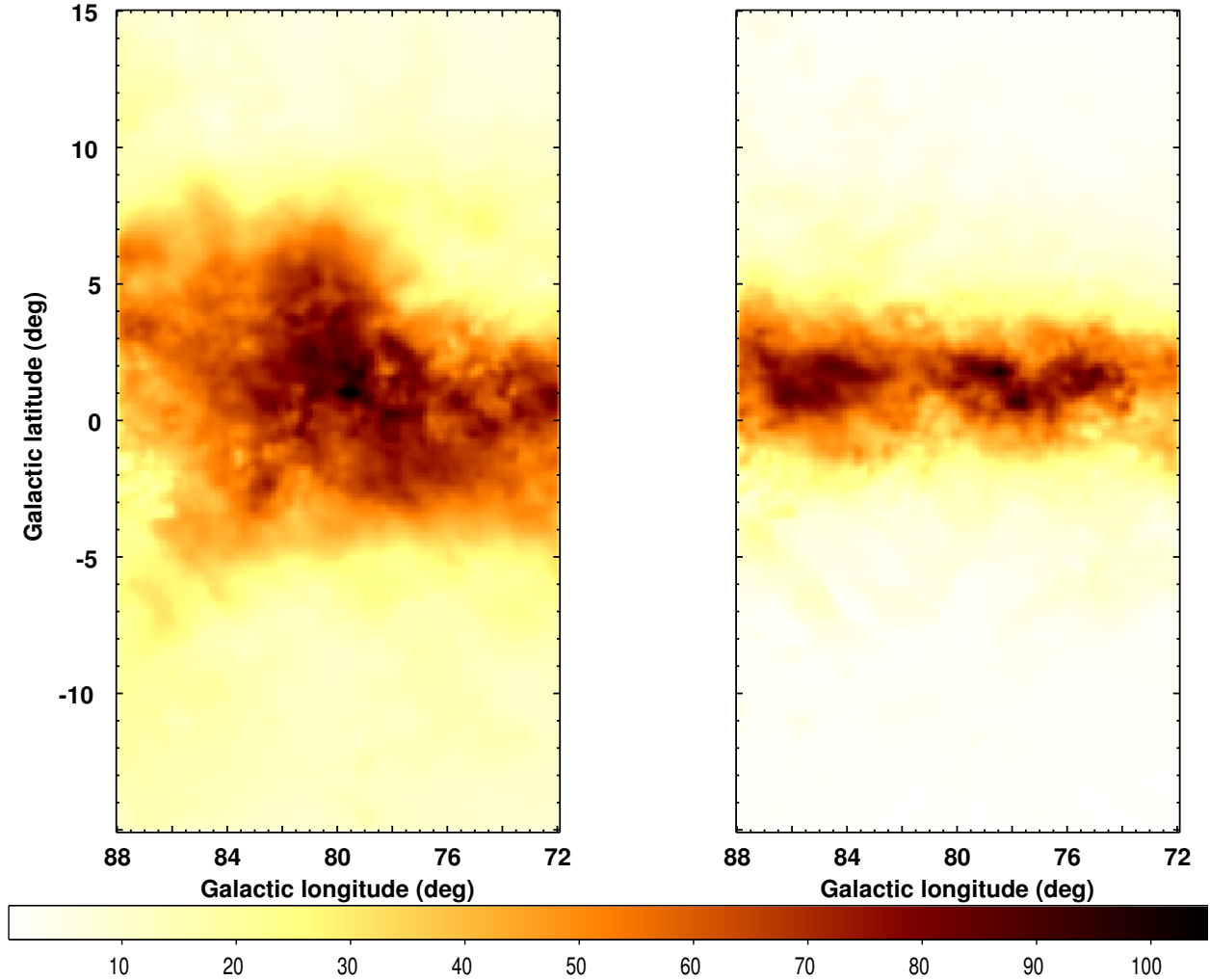}
\caption{Maps of $\nhi$ column densities in the Cygnus complex in the Local Spur (left) and in the
outer Galaxy (right), under the
assumption of a uniform spin temperature of 250~K. The color scales with $\nhi$ in units of
$10^{20}$~atoms~cm$^{-2}$.
The maps were smoothed with a Gaussian kernel of $\sigma=0.25\degr$ for display.}
\label{himaps}
\end{figure*}

\begin{figure*}
\centering
\includegraphics[width=17cm]{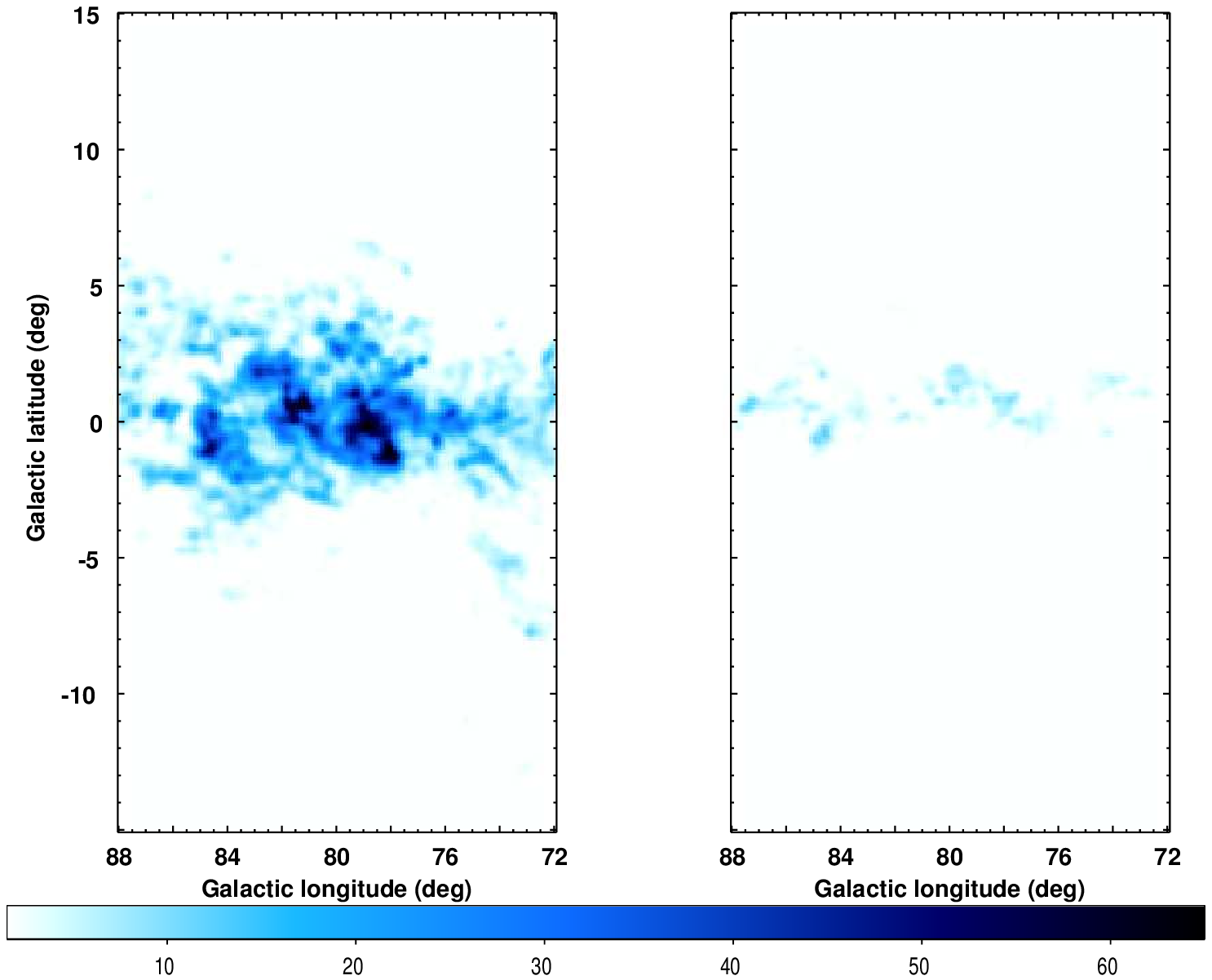}
\caption{Maps of $\wco$ intensities in the Cygnus complex in the Local Spur (left) and in the
outer
Galaxy (right). 
The color scales with $\wco$ in units of
K~km~s$^{-1}$ above 1.5~K~km~s$^{-1}$. The maps were smoothed with a Gaussian kernel of
$\sigma=0.25\degr$ for display.}
\label{comaps}
\end{figure*}

The Doppler shift of radio/mm-wave lines can be used to kinematically separate the Cygnus
{ complex}
from two faint segments of the Perseus and outer
spiral arms seen beyond Cygnus in the same direction. We applied the kinematic separation
procedure
illustrated 
by \citet{abdo2010cascep}, starting from a
preliminary boundary located at a Galactocentric radius\footnote{Based on the assumption of a flat
rotation curve with solar radius $R\sun=8.5$ kpc and Galactic rotation
velocity at the solar circle $V\sun=220$~km s$^{-1}$} $R=9.4$~kpc and then adapting
the separation
to the cloud structures and correcting for the spill-over due to the broad velocity dispersion of
\hi\ lines. The separation into two regions is
{ accurate enough} to model the interstellar \g-ray emission in Cygnus since
\citet{abdo2010cascep} and \citet{ackermann20113quad} did not find any significant
gradients of the gas \g-ray emissivities
in the outer region of the Milky Way. We applied the kinematic separation procedure to prepare maps
of the column densities of atomic gas, $\nhi$,
and of $\wco$. The maps are shown in Fig.~\ref{himaps} for \hi\ and Fig.~\ref{comaps} for CO.
All the gas maps mentioned in the paper have $> 10^\circ$ borders around the
analysis region used to
properly convolve the model with the LAT PSF.

Substantial uncertainties in the determination of $\nhi$ arise from the choice of spin temperature
for the optical
depth correction. We adopted  a uniform $T_S=250$~K as baseline case, which is
the average spin temperature that best reproduces the blending of cold and warm atomic gas according
to observations of emission-absorption $\hi$ pairs in the region covered by the CGPS
\citep{dickey2009}.
Other values $100$~K~$\leq T_S < \infty$ will be
considered later to evaluate the related systematic uncertainties affecting the results of our
analysis.

\subsubsection{Visual extinction: dark neutral gas}\label{avmaps}

Multiwavelength observations indicate that the combination of the
\hi\ and CO
lines does not properly trace the total column densities of the neutral interstellar medium (ISM)
\citep[e.g.][]{magnani2003,grenier2005,abdo2010cascep,langer2010,ade2011dg}. Since the work by
\citet{grenier2005}, dust tracers have been used in \g-ray analyses to complement the \hi\ and CO
lines, { under the assumption that dust grains are well mixed with gas in the warm and
cold phases of the ISM and therefore provide an estimate of total gas column densities}.
\citet{grenier2005} and \citet{abdo2010cascep}
adopted the $\ebv$ color excess map by \citet{schlegel1998} as a tracer of the
total column densities, and used the  $\ebv$ residuals --i.e. $\ebv$ minus the best-fit linear
combination of $\nhi$ and $\wco$ maps-- as a tracer of the dark-gas column densities in nearby
clouds
of the Gould Belt.

The use of the $\ebv$ map is problematic in the Cygnus~X region for two reasons: 
\begin{itemize}
\item numerous infrared point sources contaminate the map;
\item the temperature correction used by \citet{schlegel1998} to derive the dust column-density map
from \textit{IRAS}/ISSA measurements
is highly uncertain in regions of massive star-formation because of the enhanced radiation fields.
\end{itemize}

We have therefore adopted the visual extinction $\av$ as derived from the reddening of
near-infrared sources in the
2MASS catalog \citep{skrutskie2006}. The $\av$ maps produced by
\citet{rowles2009} and \citet{froebrich2010} were used for $\av < 5$ mag. They exhibit saturation
effects at higher extinction values, so we complemented them with an $\av$ map
obtained from 2MASS data using the code and method developed by \citet{schneider2010}. The
latter use the Besan\c{c}on stellar population model \citep{robin1986,robin2003} to
filter out the contribution from the foreground bluest stars\footnote{To do so, a distance from
the observer needs to be assumed for the clouds under consideration. We verified that variations of
a few hundred parsecs do not significantly change the results presented in the paper.}.
The second $\av$ map was built in a
$12\degr$ region centered on $(l,b)=(80\degr,0\degr)$, and, compared with the first set of maps, it
presented an offset of $\sim 0.46$ mag at low extinction. We constructed the final $\av$ map from
the
direct
\cite{rowles2009} data below 5 mag and from the second map, offset by 0.46~mag, at higher
extinction.

\begin{figure}
\centering
\resizebox{\hsize}{!}{\includegraphics{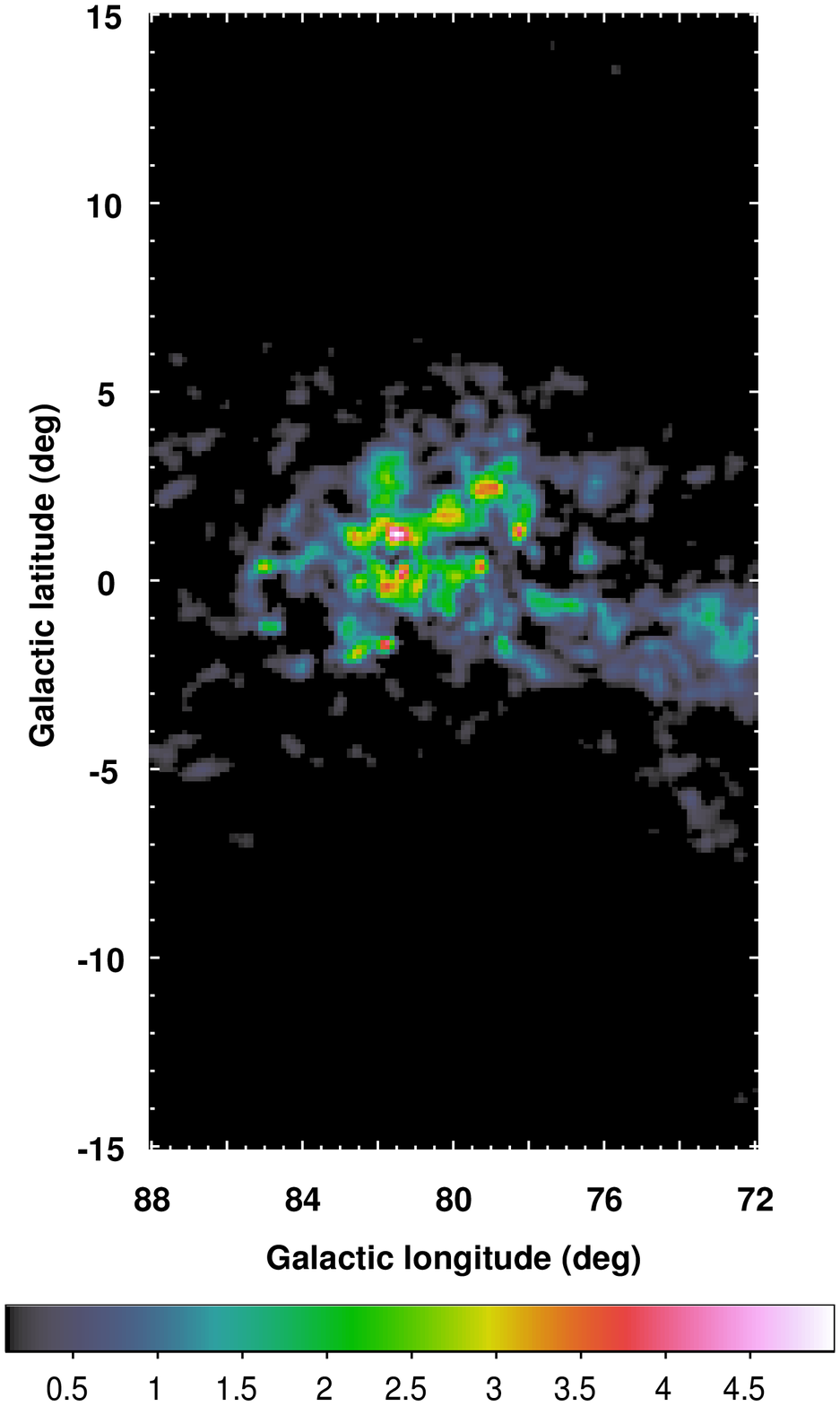}}
\caption{$\av$ excess map (magnitudes) obtained from the optical extinction $\av$ estimated
from 2MASS data minus the best-fit linear combination of the $\nhi$ and $\wco$ maps shown in
Figs.~\ref{himaps} and~\ref{comaps}, respectively. The map was smoothed using a Gaussian
kernel of $\sigma=0.25\degr$ for display.}
\label{dustmaps}
\end{figure}

The $\av$ map was binned onto the same $0.125\degr\times 0.125\degr$ grid in Cartesian
projection as the other maps. The $\av$ map was fitted with a 
linear combination of the 
$\nhi$
and $\wco$ maps previously described. The input $\av$ map minus the best-fit linear combination of
the $\nhi$
and $\wco$ maps yielded the $\av$ excess map, $\avres$, which will be used to trace the dark neutral
gas.
Only residuals corresponding to input $\av > 0.3$~mag were
kept to limit the
noise off the plane. The
$\av$ excess map is shown in Fig.~\ref{dustmaps}.

{

\subsubsection{Microwave emission: ionized gas}\label{freefreemap}
Away from $\hii$ regions around massive stars and stellar clusters,
the ionized gas constitutes a
layer of characteristic height $\gtrsim 1$~kpc over the Galactic
plane with little mass compared to the neutral phases \citep{cordes2002}. Therefore, it has often
been neglected in previous \g-ray studies.
However, we find in the Cygnus~X region many conspicuous $\hii$ regions excited by the
intense radiation fields of the numerous massive stars \citep{uyaniker2001,paladini2003}.

Ionized gas masses can be traced by free-free emission following the prescription by
\citet{sodroski1989,sodroski1997} to derive the $\nhii$ column densities:
\begin{equation}\label{Eq:nhii}
\nhii = 1.2 \times 10^{15} \mathrm{cm}^{-2} \: \left(\frac{T_e}{1\,\mathrm{K}}\right)^{0.35}\!\!
\left(\frac{n_\mathrm{eff}}{1\,\mathrm{cm}^{-3}}\right)^{-1} \!\!
\left(\frac{\nu}{1\,\mathrm{GHz}}\right)^{0.1} \!\!
\frac{I_\mathrm{ff}}{1\,\mathrm{Jy}\,\mathrm{sr}^{-1}}\mathrm{,}
\end{equation}
where $I_\mathrm{ff}$ is the free-free emission intensity at the frequency $\nu$, $T_e$ is the
electron temperature, and $n_\mathrm{eff}$ the effective electron number density. We adopted a
free-free emission map
derived from the seven-year \textit{WMAP} data in the Q~band ($40$~GHz) by \citet{gold2010} using
the maximum
entropy method from the prior template given by the extinction-corrected H$\alpha$ map by
\citet{finkbeiner2003}. It was rebinned onto the $0.125\degr \times 0.125\degr$ grid used for
the other maps, as shown in Fig.~\ref{Fig:ionmap}.
\begin{figure}
\centering
\resizebox{\hsize}{!}{\includegraphics{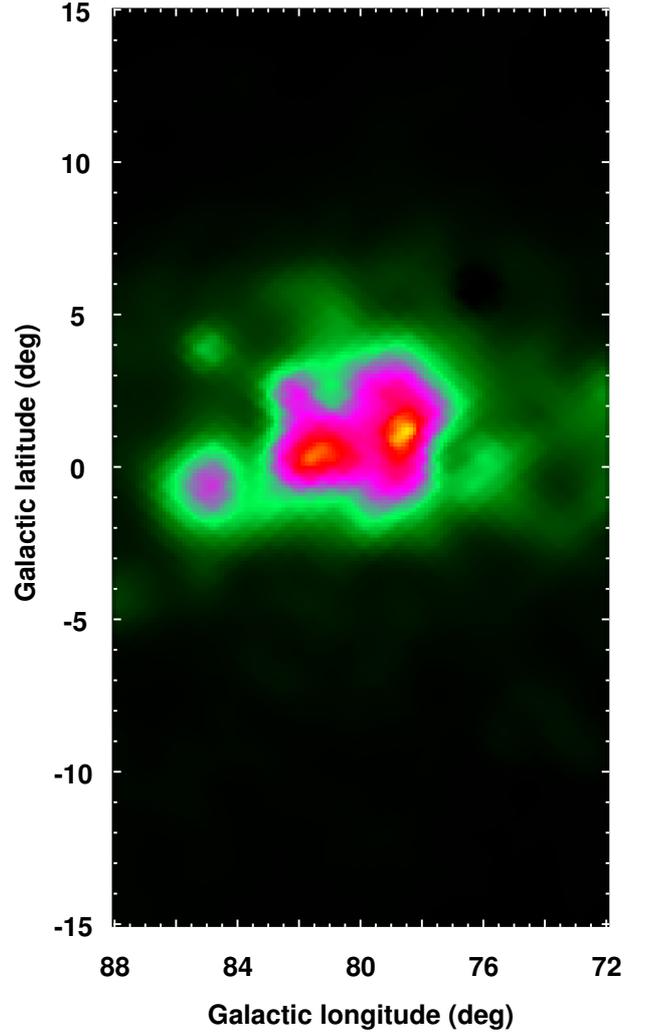}}
\caption{Free-free emission intensities from \textit{WMAP} data. The
color scales with brightness temperature in mK. The map was smoothed using a Gaussian kernel of
$\sigma=0.25\degr$ for display.}
\label{Fig:ionmap}
\end{figure}

}

\section{Analysis}

\subsection{Analysis model}

\subsubsection{Diffuse emission}\label{diffmodel}

{  Since the bulk of Galactic CRs in the relevant energy ranges are expected to be
smoothly distributed on scales }
exceeding the typical dimensions of interstellar clouds and to penetrate all
the phases of the ISM uniformly, the \g-ray emission produced by CR-gas interactions can be
modeled to first
order as a linear
combination of the gas column densities summed for the different phases and
different regions along the line of sight.

{
Ionized gas, with a total mass over the region of $0.4 \times 10^6\, 
(n_\mathrm{eff}/1\,\mathrm{cm}^{-3})^{-1} \, M\sun $ for $T_e=10^4$~K, represents less
than $4\%$ of the total atomic mass present in the Cygnus complex (assuming
$n_\mathrm{eff}=2-10$~cm$^{-3}$, \citealt{sodroski1997}).
The corresponding column densities are highest in the massive star-forming
region of
Cygnus X where we detected a bright and hard extended \g-ray source powered by freshly-accelerated
particles, that will be
called hereinafter ``the cocoon'' \citep{cocoonScience}. The free-free
emission map was significantly detected in addition to the
other interstellar components, but only at the expense of an unusually hard spectrum. 
To model the entire region, we introduced an extended source to account for the
cocoon, as described in~\ref{srcmodel}. The latter was found to provide the best fit to the
LAT
data, yielding a higher maximum-likelihood value than the free-free emission template.
Since the cocoon source overlaps most of the ionized clouds, it absorbs their
contribution to the \g-ray emission, so ionized gas was not
included in the baseline model through the free-free emission template. The Cygnus X region is
treated in detail in a
companion paper \citep{cocoonScience}.
The results presented in this paper
were checked against the inclusion of the free-free emission template in the model.} 

The interstellar IC emission is produced by interactions of CR electrons and
positrons with the low-energy interstellar radiation field (ISRF). To account for large-scale
IC emission from the Milky Way we adopted a
template
calculated using the GALPROP CR
propagation code\footnote{http://galprop.stanford.edu}
\citep{strong1998,strong2007}, run \texttt{54\_87Xexph7S}. The IC
emission was calculated on the basis of a CR
electron spectrum consistent with recent measurements { at the Earth}
\citep{abdo2009eposspec} and
the new calculation of the Galactic ISRF by \citet{porter2008}.

Local
radiation fields could leave unmodeled structures in IC emission,
{ notably in the massive star-forming region of Cygnus X \citep[e.g.][]{orlando2007}. In
the companion paper we show that an upper bound to the IC emission from the stellar and
interstellar low-energy radiation fields upscattered by CR electrons with the local spectrum is two
orders of magnitude fainter than the cocoon emission, which in turn is fainter than the emission
from
the neutral gas (Fig.~\ref{cumspec}). The CR
electron sources within Cygnus X
could further enhance the IC \g-ray yield. Any enhanced IC
contribution from the inner region is accounted for in this analysis by the extended cocoon
source, and it should not bias
the
determination of the gas emissivities we aim to study here.}

The diffuse emission model is completed by the isotropic
background which combines the residual backgrounds from misclassified
CR interactions in the LAT
and the isotropic, presumably extragalactic, \g-ray emission \citep[studied in detail
in][]{abdo2010isoback}.

\subsubsection{Sources}\label{srcmodel}

We included in the model the identified sources in the region of interest: Cygnus X-3
\citep{abdo2009cygx3}, PSR
J1957$+$5033 \citep{sazparkinson2010} and PSR
J2030$+$3641 \citep{camilo2011}, in addition to the three bright pulsars as discussed
in \S~\ref{psr}.

We also iteratively included significant 1FGL point sources \citep{abdo20101FGL}
either associated with active galactic nuclei (AGN) or characterized by variability or both.
The sources were added with
decreasing brightness:
J2116.1$+$3338, J2001.1$+$4351, J2027.6$+$3335,
J2115.5$+$2937, J2015.7$+$3708, J2029.2$+$4924, 
J2012.2$+$4629, and J2128.0$+$3623.
The iterative procedure is useful for stabilizing the likelihood fitting
procedure and assessing the significance of sources added at each step.

We detected extended \g-ray emission { above the global interstellar emission model
discussed here} associated with the supernova
remnants known as the Cygnus Loop \citep[G74.0-8.5, e.g.][]{sun2006} and \g~Cygni
\citep[G78.2+2.1, e.g.][]{ladouceur2008} and with the inner 100~pc of the Cygnus X complex.
They are discussed in detail elsewhere.
We briefly
summarize here how these extended sources are modeled. For each of them we have tested
different
models and therefore verified that
their presence does not bias the results concerning the properties of large-scale interstellar
emission presented in the paper.

The Cygnus Loop was modeled using a ring centered at $(l,b)=(74.1\degr,-8.5\degr)$
and with inner/outer radii of $0.7\degr$ and $1.6\degr$, respectively, which best reproduces \g-ray
emission from the Cygnus Loop { \citep{katagiri2011}}.

We included two sources in the region of \g~Cygni in addition to PSR J2021$+$4026:
\begin{itemize}
 \item a uniform disk centered at $(l,b)=(78.2\degr,+2.1\degr)$ and a radius of $0.5\degr$
\citep[G78.2$+$2.1;][]{green2009};
 \item a 2D Gaussian corresponding to the moderately extended TeV source\footnote{The 2D Gaussian
fitted to TeV data by \citet{weinstein2009} provides a better fit to LAT data with respect to
the coincident point source 1FGL~J2020$+$4049.} VER~2019$+$407
\citep{weinstein2009}.
\end{itemize}

\begin{figure}
\resizebox{\hsize}{!}{\includegraphics{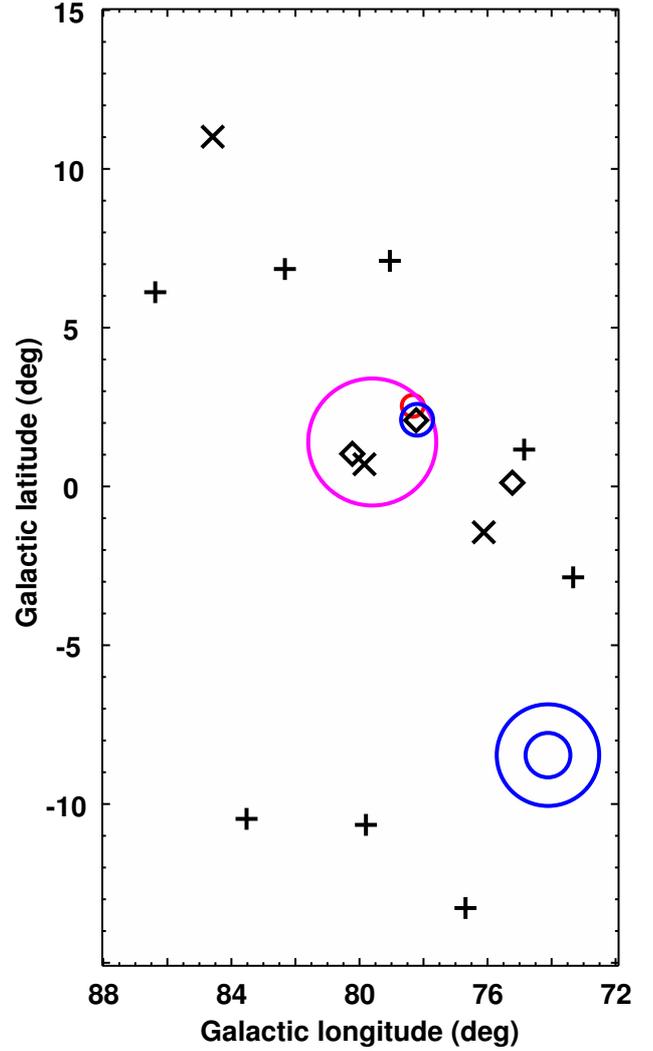}}
\caption{Sources included in the analysis model. Diamonds mark the positions of the three
bright pulsars { that were dimmed by phase selection} (\S~\ref{psr}). X points mark
the positions
of other identified sources. Crosses correspond to 1FGL sources either associated to AGN or
variable, or both, confirmed by our analysis. The blue circles correspond to the rims of the
templates adopted to model SNRs, the Cygnus Loop (G74.0$-$8.5) and \g~Cygni (G78.2$+$2.1). The red
circle marks the centroid of VER~2019$+$407 (whose extension is not appreciable in this large-scale
view). The magenta circle represents the $1\sigma$ contour of the Gaussian used to model the
cocoon.}
\label{finderchart}
\end{figure}

We detected extended \g-ray emission toward the inner $\sim 100$~pc of Cygnus~X, { which
is effectively treated here as a source named ``the
cocoon''}. We discuss the nature of the cocoon { and
its relation with CR acceleration in the massive star-forming region} in
a
dedicated paper \citep{cocoonScience}, where we determine a Gaussian
centered at $(l,b)=(79.6\degr \pm 0.3\degr,1.4\degr \pm 0.4\degr)$ with
a $\sigma=2.0\degr \pm 0.2\degr$ width to be the model providing the best fit to the LAT data.
{ As noted above, the cocoon source effectively accounts for the contribution from ionized
gas and from enhanced IC emission in the Cygnus X massive star-forming region, as well as
for locally-accelerated CRs.} The spatial distribution of the sources included in the analysis model
is summarized in
Fig.~\ref{finderchart}.

\subsubsection{Summary of the analysis model}

To summarize, the \g-ray intensities
$I$
(photons cm$^{-2}$ s$^{-1}$
sr$^{-1}$) are modeled in each energy bin by 
\begin{eqnarray}
I(l,b) & = & \sum_{\imath=1}^2 \left[ \qhi{\imath} \cdot \nhi(l,b)_\mathrm{\imath}+
\qco{\imath} \cdot \wco(l,b)_\mathrm{\imath}\right]+ \nonumber\\
&+& \qdust \cdot \avres (l,b) + \mathrm{IC}(l,b) + I_\mathrm{iso} + \nonumber\\
&+& \label{gasmodel} \sum_\jmath S_\jmath(l,b).
\end{eqnarray}
The sum over $\imath$ represents the combination of the two regions: 1)~Cygnus complex and
2)~outer
Galaxy.
The free parameters of the diffuse emission model are the emissivities per hydrogen atom,
$\qhi{\imath}$
(s$^{-1}$ sr$^{-1}$), the
emissivities per unit of $\wco$ intensity, $\qco{\imath}$ (cm$^{-2}$ s$^{-1}$ sr$^{-1}$ (K km
s$^{-1}$)$^{-1}$), the emissivity per $\av$ excess unit $\qdust$ (cm$^{-2}$ s$^{-1}$
sr$^{-1}$ mag$^{-1}$), and the isotropic intensity $I_\mathrm{iso}$ (cm$^{-2}$ s$^{-1}$
sr$^{-1}$). $\mathrm{IC}$ stands for { the IC emission model described above}. The sum
over
$\jmath$ represents the combination of the sources, either point-like or
extended, as described in \S~\ref{srcmodel}, including a free parameter (flux normalization)
independently for
each of them.                                                                                      

\subsection{Analysis method}\label{anamethod}
The model was fit to LAT data by using a binned
maximum likelihood with
Poisson statistics\footnote{As implemented in the standard LAT analysis tools \texttt{09-18-05}.}
independently over several energy bins.
We
used a $0.125\degr \times 0.125\degr$ binning in Cartesian projection, comparable to
the LAT angular
resolution at the highest energies. We considered three energy bands: low
($100$ MeV--$1$ GeV), mid ($1$ GeV--$10$ GeV) and high energies ($10$ GeV--$100$ GeV). The low and
mid-energy
bands were divided further into four logarithmic-spaced energy bins, the higher-energy band in
two because of
the limited statistics\footnote{The bounds of the energy bins are reported in
Table~\ref{fitparam}.}. The analysis was based on the post-launch IRFs of
the \texttt{P6\_V3} series,
which consider efficiency losses due to pile-up and accidental coincidence effects in the
detector \citep{rando2009}.

To perform the convolution with the LAT PSF, a power-law spectrum with index $2.1$ was assumed
for the
gas maps and other sources modeled by geometrical templates { (the
results do not significantly depend
on this value)}. For all other sources we
used power-law spectra with the
spectral index reported in the 1FGL Catalog.  For
the pulsars
included in the LAT pulsar catalog \citep{abdo2010psrcat}, we used the spectral
functions
described therein.

\section{Results and discussion}

\subsection{Summary of the results and uncertainties}
\begin{figure*}
\centering
\includegraphics[width=17cm]{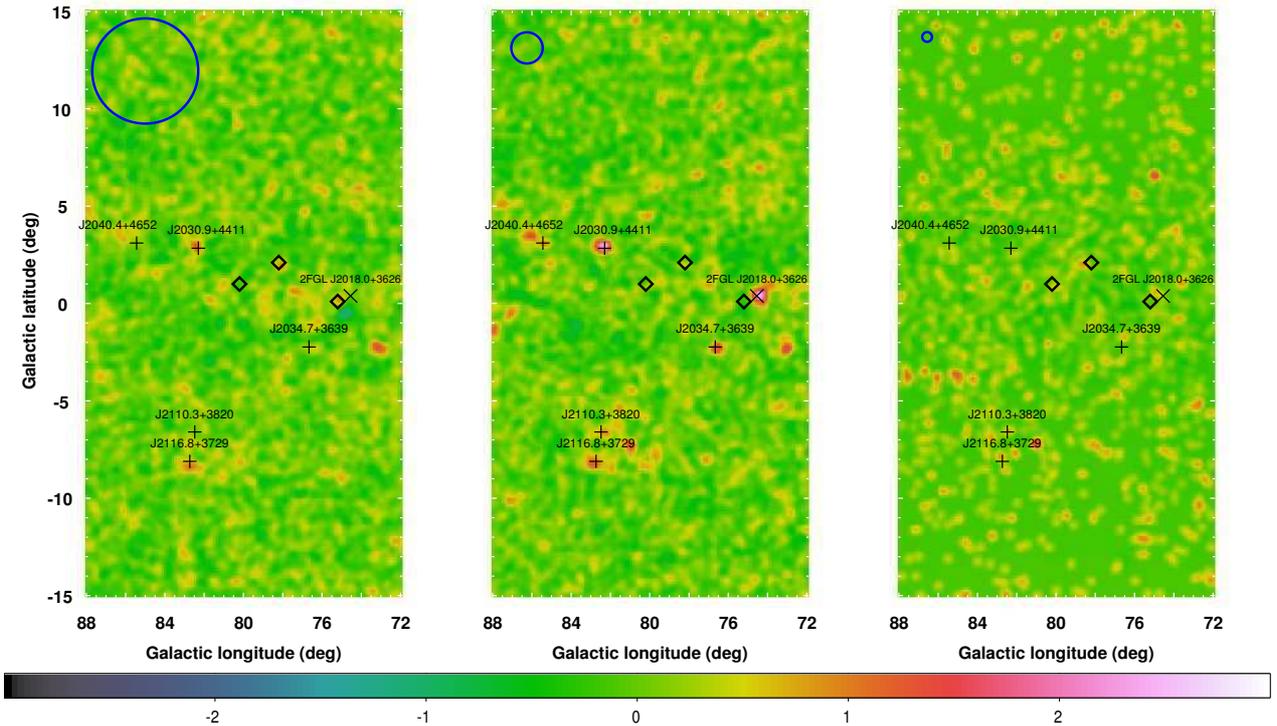}
\caption{Residuals (data--model). Left: low energies ($100$~MeV--$1$~GeV); center: mid-energies
($1$~GeV--$10$~GeV); right: high energies ($10$~GeV--$100$~GeV). Units are approximate standard
deviations
(square
root of model counts) saturated between
$\pm 3$ and smoothed for display with a Gaussian kernel of
$\sigma=0.25\degr$. { In each panel the blue circle in the top left corner represents the
effective LAT PSF~68\% containment circle for the event selection used in the analysis (averaged
over the corresponding
energy range assuming a power-law spectrum with index 2.1).} Diamonds mark the
positions of bright pulsars for which phase selection was
applied, as in Fig.~\ref{countmap}. Crosses mark the positions of unassociated 1FGL sources
coincident with positive residuals; the X point mark the position of
2FGL~J$2018.0+3626$, coincident with a hot spot in the residual map $1$
~GeV--$10$~GeV.}
\label{resid5c}
\end{figure*}

The \g-ray residuals corresponding to the best-fit model are shown in Fig.~\ref{resid5c}.
They indicate that the model satisfactorily reproduces
the morphology of the \g-ray emission on larger angular scales than the LAT PSF in all
the energy bands.
Localized positive
residuals
are still present. Some of them coincide with unassociated 1FGL sources and others are associated
with
sources of the 2FGL catalog\footnote{The preparation of the 2FGL catalog ran in parallel with the
analysis reported in this paper. The source list is now available from
{ http://fermi.gsfc.nasa.gov/ssc/data/access/lat/2yr\_catalog/}.}, notably
2FGL~J$2018.0+3626$ also coincident with the TeV source MGRO~J2019+27 \citep{abdo2007}. We verified
that including sources accounting for those residuals in the analysis model would not significantly
affect the determination of \g-ray emissivities associated with the different gas
components summarized in Table~\ref{fitparam}.
\begin{table*}[tbp]
 \caption{\label{fitparam}Best-fit parameters describing
emission from interstellar gas (Eq.~\ref{gasmodel}) under the assumption of a uniform \hi\ spin
temperature $T_S=250$~K.}
\centering
\begin{tabular}{lr@{$\pm$}lr@{$\pm$}lr@{$\pm$}lr@{$\pm$}lr@{$\pm$}l}
\hline\hline
energy bin\tablefootmark{a}&
\multicolumn{2}{c}{$\qhi{1}$}\tablefootmark{b}&
\multicolumn{2}{c}{$\qco{1}$}\tablefootmark{c}&
\multicolumn{2}{c}{$\qhi{2}$}\tablefootmark{b}&
\multicolumn{2}{c}{$\qco{2}$}\tablefootmark{c}&
\multicolumn{2}{c}{$\qdust$}\tablefootmark{d}\\
\hline
0.1--0.178	& 7.9&1.1	& 3.3&1.0	& 8.5&1.4	& 0.00&0.06	& 10&30\\
0.178--0.316	& 5.9&0.3	& 2.2&0.3	& 4.7&0.4	& 0.000&0.002	& 19&5\\
0.316--0.562	& 3.27&0.11	& 1.16&0.08	& 3.23&0.14	& 0.00&0.06	& 11.0&1.8\\
0.562--1	& 1.95&0.06	& 0.59&0.04	& 1.72&0.10	& 0.5&0.3	& 6.2&0.7\\
1--1.78		& 0.98&0.03	& 0.328&0.016	& 0.74&0.04	& 0.12&0.13	& 2.6&0.3\\
1.78--3.16	& 0.389&0.016	& 0.141&0.008	& 0.36&0.02	& 0.02&0.06	&0.86&0.15\\
3.16--5.62	& 0.151&0.005 	& 0.044&0.004	& 0.113&0.013	& 0.02&0.03	&0.39&0.08\\
5.62--10	& 0.050&0.003	& 0.016&0.002	& 0.046&0.006	& 0.000&0.005	&0.15&0.04\\
10--31.6	& 0.0085&0.0015	& 0.0059&0.0010	& 0.021&0.003	& 0.002&0.008	&0.043&0.019\\
31.6-100	& 0.0024&0.0007	& 0.0016&0.0004	& 0.0007&0.0014	& 0.002&0.003	&0.002&0.007\\
\hline
\end{tabular}
\tablefoot{Subscripts refer to the two regions separated in analysis: 1) the Cygnus complex in
the Local Spur, 2) the outer Galaxy. Some parameters are poorly determined but
they are reported for completeness.\\
\tablefoottext{a}{GeV}
\tablefoottext{b}{$10^{-27}$ s$^{-1}$ sr$^{-1}$}
\tablefoottext{c}{$10^{-6}$ cm$^{-2}$ s$^{-1}$ sr$^{-1}$ (K km s$^{-1}$)$^{-1}$}
\tablefoottext{d}{$10^{-6}$ cm$^{-2}$ s$^{-1}$ sr$^{-1}$ mag$^{-1}$}}
\end{table*}

\begin{figure*}
\centering
\includegraphics[angle=90,width=17cm]{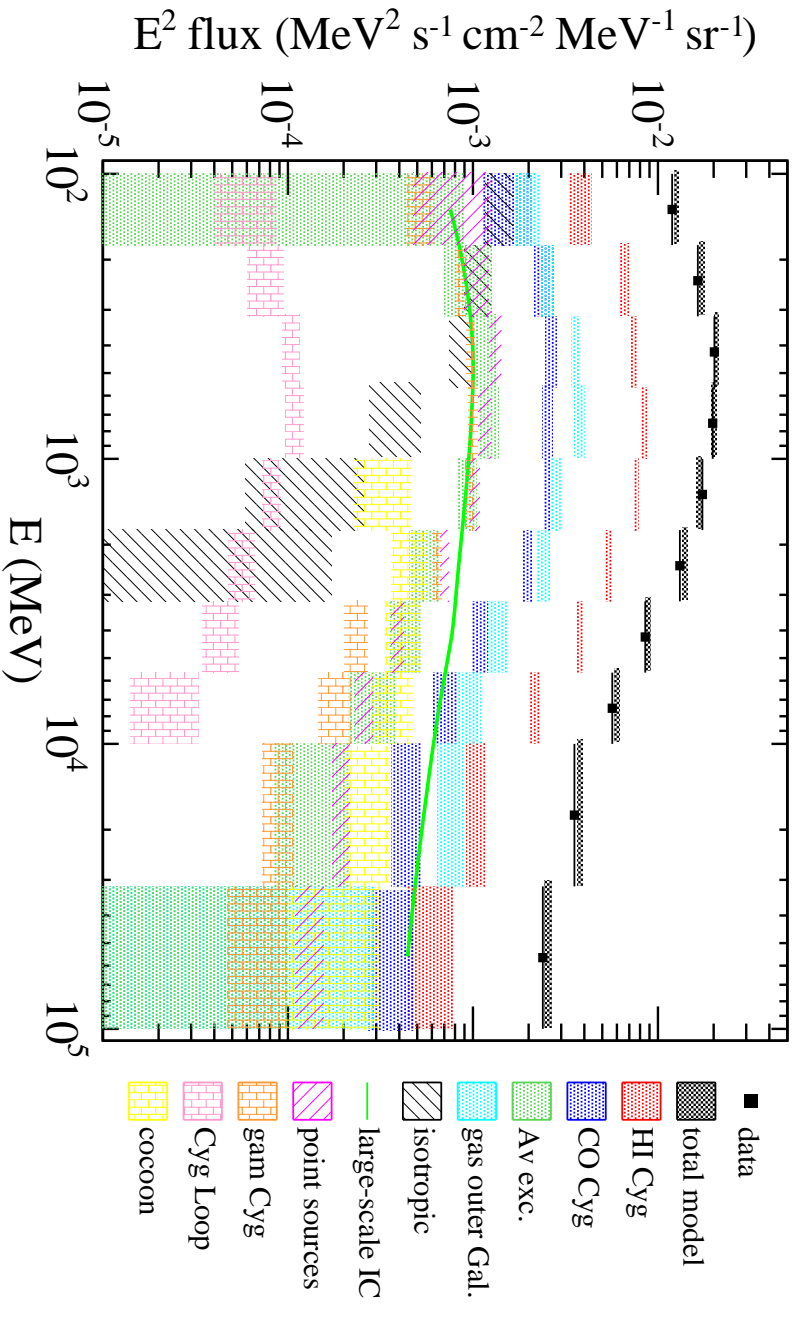}
\caption{Spectral energy distribution of \g-ray emission measured by the LAT
compared with our best-fit model. { The uncertainties shown are statistical only}. We
separately show
the different
components of the interstellar emission model, point sources and extended objects. The curve
corresponding to \g~Cygni combines the contributions of the off-pulse source, the disk associated
with the remnant and the 2D Gaussian accounting for VER~2019$+$407.}
\label{cumspec}
\end{figure*}

Figure~\ref{cumspec} shows the \g-ray spectral energy distribution measured by the LAT over the
whole
region of interest.
The LAT measurements are compared with the final model, and the different components are outlined.
The data sample is dominated by emission from interstellar gas in the Cygnus
complex. The largest contributor is $\hi$. Emission associated with CO and $\av$
excesses exceeds the signals from individual sources for the whole energy
range considered. The cocoon has a
very hard spectrum and becomes comparable to emission from CO-bright gas at energies $>10$~GeV.

\begin{figure}
\centering
\resizebox{\hsize}{!}{\includegraphics{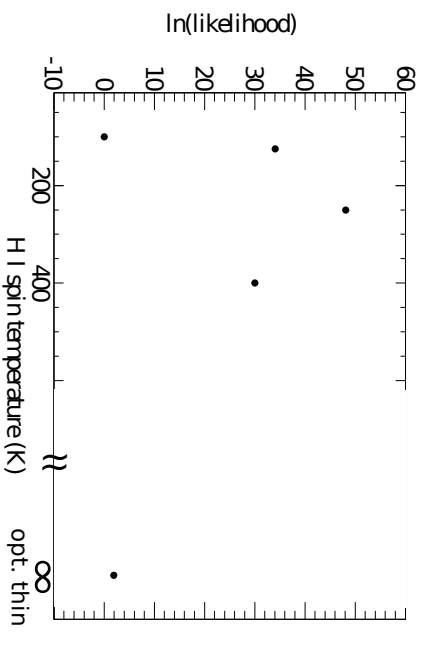}}
\caption{Maximum likelihood obtained as a function of the uniform spin
temperature adopted for the
optical depth
correction of $\hi$\ data. Values are offset so that log-likelihood is zero for a
spin temperature of $100$~K.}
\label{TSL}
\end{figure}

All the results presented so far have been based
on the assumption of a uniform $\hi$\ spin temperature of $250$~K. To gauge the impact of the
optical depth correction of $\hi$\ data on the
results, we repeated the analysis with other assumptions.  $T_S=400$~K is considered since it is
the value best reproducing pairs of emission/absorption $\hi$ spectra over most of the regions
analyzed by \citet{dickey2009}, although
they find that $T_S=250$~K is preferred in the region covered by CGPS data.
$T_S=125$~K is considered because it has long been used for \g-ray studies
\citep[e.g.][]{bloemen1984,hunter1997,strong2004}. We also considered
two extreme assumptions: a low\footnote{The spin temperature is higher than the
brightness temperature, measured $>100$~K along many directions in the region.}
$T_S=100$~K,
and the optically thin approximation (equivalent to infinitely high spin temperature).
Figure~\ref{TSL}
shows the maximum likelihood profile obtained for the final model as a function of $T_S$.
The results support the average spin
temperatures of a few hundred~K deduced from
radio absorption/emission measurements by \citet{dickey2009}, implying a mix of $<
25\%$ cold and $> 75\%$ warm $\hi$.

{ 
To test the robustness of the results against the presence of ionized gas beyond the
extended cocoon source, we replaced the latter in the baseline model with the free-free emission
template. All the results for the gas emissivities were found to be consistent with the
values listed in Table~\ref{fitparam} within statistical
errors. 
}

The large-scale IC model introduced in \S~\ref{diffmodel} is affected by considerable
uncertainties related to the distribution of CR densities and of the ISRF in the Galaxy. The
intensity of IC emission expected over our region of interest is comparable to the one
from interstellar gas. The latter, however, is highly structured and has a lower characteristic
height above the Galactic plane, and can therefore be reliably
determined in the likelihood fit. We verified that completely neglecting the large-scale IC
emission leads to negligible variations in the emissivities of CO-bright gas and $\av$ excesses,
and to variations lower than or comparable to statistical uncertainties for
the emissivities of the atomic hydrogen, which is less structured than the other ISM components
and has a larger characteristic height.
 
Other systematic uncertainties are due to the LAT instrument response. The
uncertainties in the \g-ray selection efficiency are estimated to be
$10\%$ at $100$~MeV, $5\%$ at $560$~MeV, 
and $20\%$ above $10$~GeV for the IRFs we used here
\citep{abdo2010isoback}. The Monte Carlo-based PSF used for this study is known
to not accurately reproduce in-flight data over the whole energy
range considered. We verified by means of
dedicated simulations that this does
not significantly affect the determination of the
gas emissivities considered for the discussion. The energy dispersion is routinely neglected
in the
likelihood fitting of LAT data for limitations in computing power: Monte Carlo simulations
indicate that this approximation causes a bias on the order of $10\%$ at 100~MeV decreasing to
$< 5\%$ above $200$~MeV.

\subsection{\hi\ emissivity and CR densities}\label{hiemiss}

The $\hi$\ emissivity per hydrogen atom relates to the average CR
density in each of the regions considered.  LAT measurements
\citep{abdo2009lochiemiss} show
that the $\hi$ emissivity spectrum in the local ISM is consistent with production
via electron Bremsstrahlung and nucleon-nucleon interactions by CRs with a spectrum
consistent with that directly measured in the neighborhood of the Earth.

\begin{figure}
\centering
\resizebox{\hsize}{!}{\includegraphics{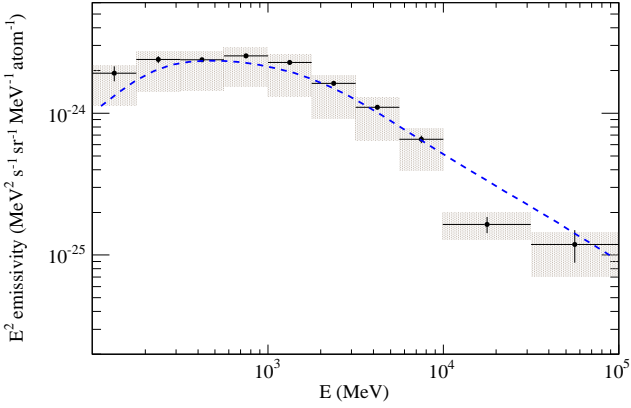}}
\caption{$\hi$\ emissivity spectrum in the Cygnus complex.
Points: the best-fit estimate for the
spin temperature $T_S=250$~K. Hatched rectangles: systematic uncertainties taking $\hi$
opacity and \g-ray selection efficiency into account. Line: model of the
local interstellar spectrum by \citealt{abdo2009lochiemiss} (with a nuclear enhancement factor of
$1.84$, \citealt{mori2009}).}
\label{hiemissplot}
\end{figure}

The integrated \g-ray emissivity $>100$~MeV we measure in the Cygnus complex amounts to
$[2.06 \pm 0.11
\,(\mathrm{stat.})\; ^{+0.15}
_{-0.84}\,(\mathrm{syst.})]\times 10^{-26}$ s$^{-1}$ sr$^{-1}$. Figure~\ref{hiemissplot} shows the
$\hi$
emissivity spectrum obtained for the Cygnus complex
and compares it with the expectations for the local interstellar spectrum estimated in
\citet{abdo2009lochiemiss}. The latter is compatible (within $10\%$) with LAT observations at mid
latitudes in the
third
Galactic quadrant in the energy range $100$~MeV--$10$~GeV \citep{abdo2009lochiemiss}.
The spectrum is presented
for a uniform spin temperature $T_S=250$~K;
systematic uncertainties due to the $\hi$\ opacity correction and to the \g-ray
selection efficiency are added in quadrature for display. The latter give a non-negligible
contribution only at energies
larger than a few GeV.

The emissivity of atomic gas in the Cygnus region, averaged over $\sim 400$ pc, 
is { consistent with the local emissivity in the
$100$~MeV--$100$~GeV energy
range,
except for the deviant point at $10-30$~GeV. This can be explained by the difficulty
distinguishing the $\hi$\ and CO components at high energies, for which the \g-ray statistics are
limited and the hard cocoon source is brighter (Fig.~\ref{cumspec})}. The
emissivity
spectrum implies that the CR spectra in the relevant energy ranges
($\sim 1-100$ GeV/n for nucleons) are similar to those
measured 
in the vicinity of the Earth and inferred from \g-ray observations in the nearby interstellar space
within $1$~kpc.

The variations in average CR densities along the Local Spur between
the dense Cygnus complex, two segments in the second and third quadrants that exhibit $\sim 2$ lower
surface densities of gas \citep{abdo2010cascep,ackermann20113quad}, and the mid-latitude diffuse
medium with a factor $\sim 5$ lower surface density \citep{abdo2009lochiemiss} are
constrained to within 10\% to 35\%. This is difficult to
reconcile with the idea of a dynamical coupling between
gas and CR densities \citep[e.g.][]{bertsch1993,hunter1997}. They are consistent, on the other
hand, with the small arm-interarm emissivity contrast estimated from LAT data in the
third Galactic quadrant \citep{ackermann20113quad}.
In spite of the high column densities of gas, exceeding $10^{22}$ atoms
cm$^{-2}$ over many directions within the Cygnus complex, we find no hints of excluding
CRs from the densest parts of the
atomic clouds.

Owing to the bright foreground of the Cygnus complex and individual sources, studying
the gas emissivity in
the outer disk of the Milky Way  in detail is beyond the scope of this study. However, the ratio
of
the integrated $\hi$\ emissivity of the outer region over that in the Local Spur is $(90\pm7)\%$, in
very good agreement with the results by \citet{abdo2010cascep} and \citet{ackermann20113quad}. It
confirms in
another direction the presence of high CR densities beyond the solar circle.

{ Located at a distance of $\sim 1.4$~kpc and $l=80\degr$, the Cygnus complex lies
at $R\simeq8.4$~kpc
from the Galactic center, at a slightly smaller radius than the solar system. 
We measure in this direction an emissivity that is
consistent with other values found in the Local Spur and in three outer segments of the
Milky Way with Galactocentric radii up to $\sim 15$~kpc. It yields a decrease in
\hi\ emissivity $< 60\%$ over $\sim 6$~kpc in Galactocentric radius across and beyond the solar
circle. Nevertheless, those measurements span a narrow range in azimuth around the Galactic center
and may
not be representative enough for comparison with axisymmetric propagation models  in order
to study the CR gradient across the solar circle .}

\subsection{CO-bright molecular gas}

\begin{figure}
\centering
\begin{tabular}{c}
\resizebox{\hsize}{!}{\includegraphics{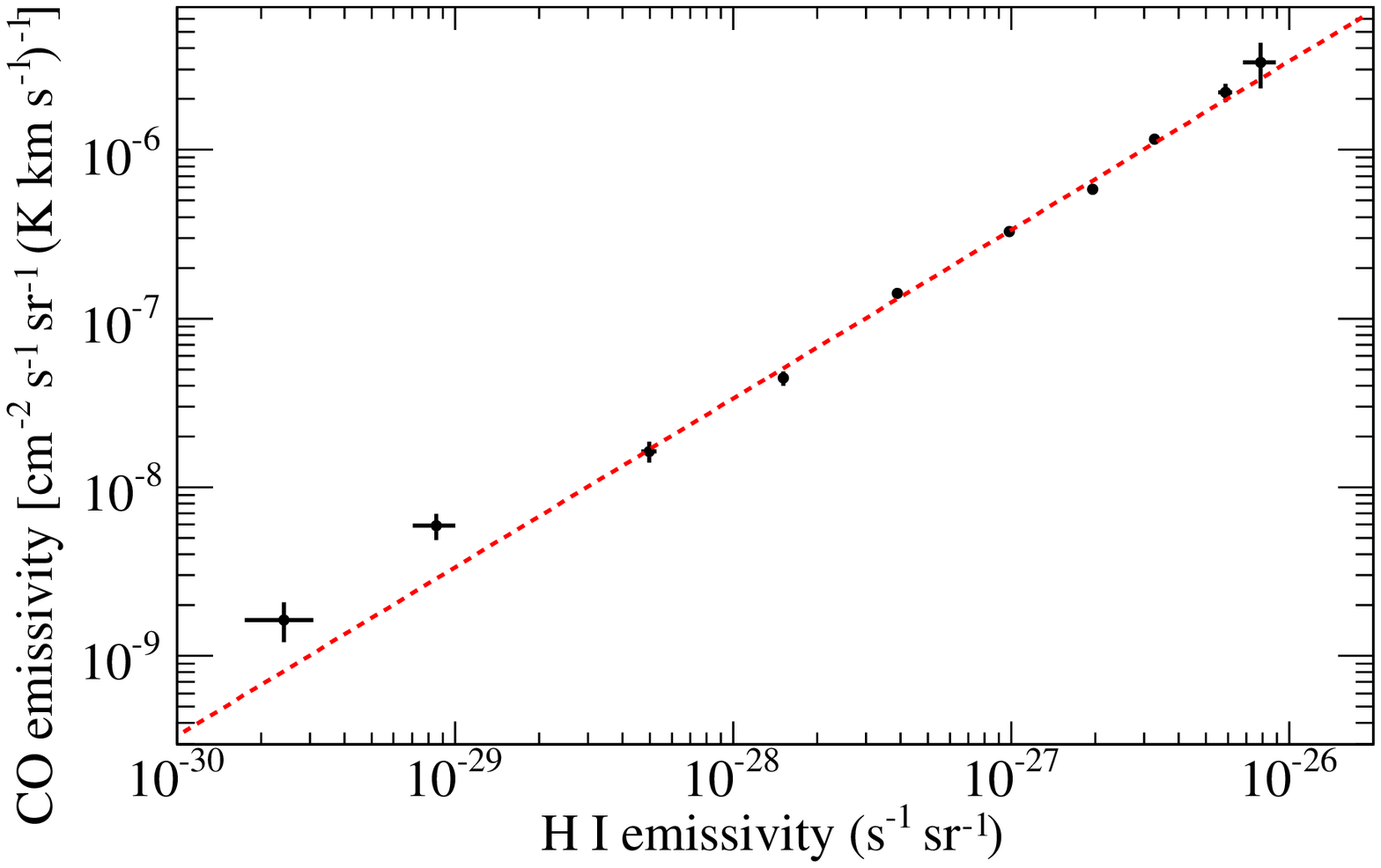}}\\
\resizebox{\hsize}{!}{\includegraphics{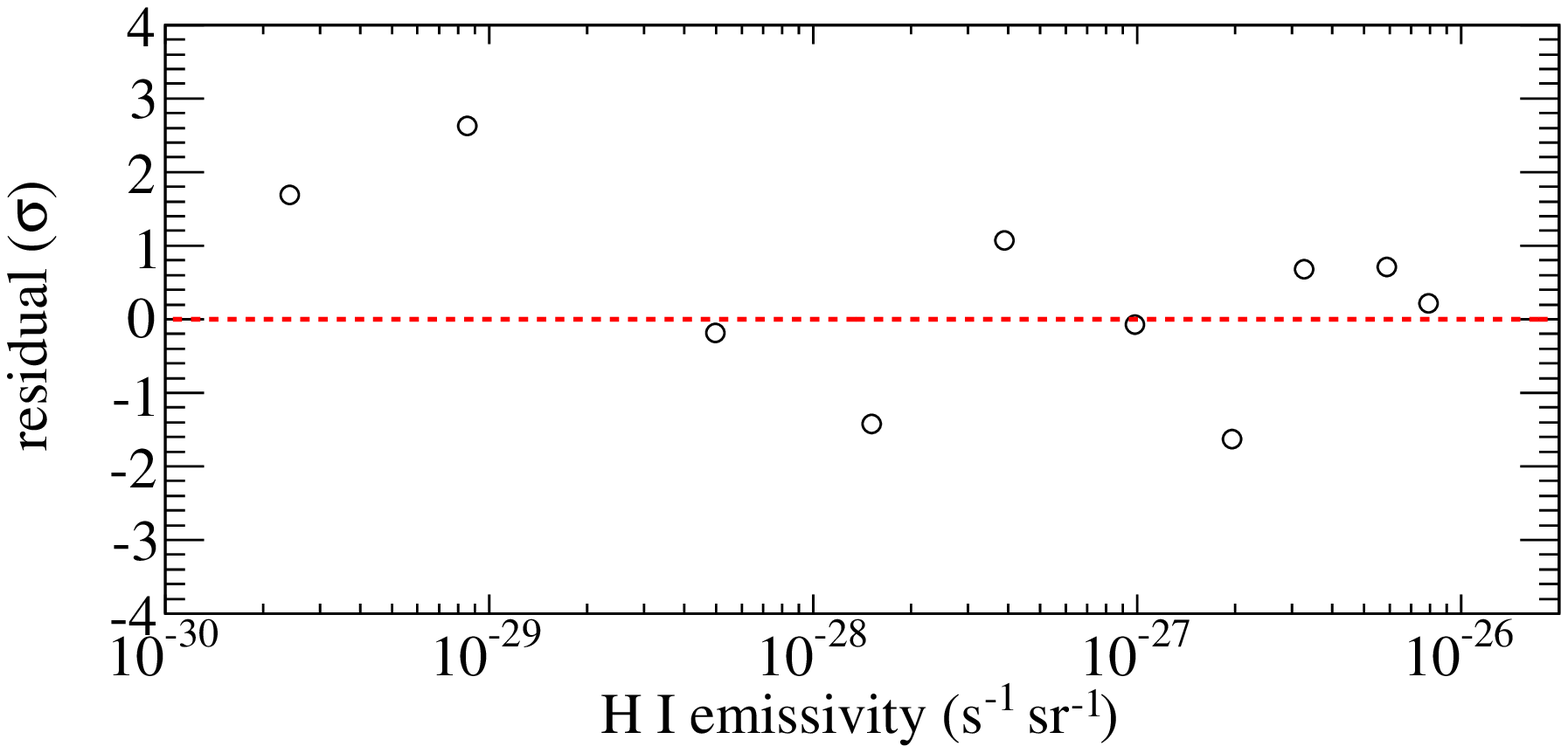}}
\end{tabular}
\caption{Top: emissivity per $\wco$ intensity unit versus emissivity per hydrogen atom in the Cygnus
complex (for $T_S=250$~K). The points correspond to the different energy bins; the emissivities
decrease with increasing energy. 
The red line gives the best linear fit taking uncertainties on both axes into account. 
Bottom: residuals in units of standard deviations as a function of $\hi$\ emissivity.}
\label{coplots}
\end{figure}

If molecular and atomic gas are illuminated by the same CR fluxes, we expect the emissivity per
hydrogen
molecule to be twice the emissivity per hydrogen atom, so we can
calibrate the
$\xco$ ratio. We performed a linear fit, $\qco{\imath}= \overline{q}+2\xco \cdot \qhi{\imath}$,
taking
the uncertainties on both emissivities into account to derive
the best linear relation shown in Figure~\ref{coplots}. We also give the residuals in units of
standard deviations.
A good linearity is found in the 0.1-10 GeV energy range. The highest energy
(lowest emissivity) points show $<3\sigma$ excess of emission associated with CO with
respect to the
$\xco$ ratio determined at low energies. The high CO emissivity recorded at $10-30$~GeV (second
point) corresponds
to
a low emissivity in $\hi$\
(Fig.~\ref{hiemissplot}) and may result from a fluctuation in the difficult spatial separation
between
the atomic and molecular components when photons are sparse { given the} hard
$2^\circ$
source that partially overlaps the CO peaks.
Up to $10$ GeV, the linearity is good, so there is no sign of CR exclusion from the
dense cores of this giant molecular complex.

The slope of the best-fit linear relation provides a value of $\xco=(1.68\pm0.05)\times 10^{20}$
cm$^{-2}$ (K km s$^{-1}$)$^{-1}$ in the case of $T_S=250$~K. We obtain $\xco=(1.58\pm0.04)\times
10^{20}$
cm$^{-2}$ (K km s$^{-1}$)$^{-1}$ in the limit of small $\hi$\ optical depth and
$\xco=(2.55\pm0.08)\times 10^{20}$
cm$^{-2}$ (K km s$^{-1}$)$^{-1}$ in the case of $T_S=100$~K. The uncertainties in $\qhinull$
associated with the
$\hi$\ spin temperature are particularly severe for the high-density clouds of the Cygnus complex.
High optical depths 
(low spin temperatures) imply a large increase in $\nhi$, therefore substantially lower CR
densities. Given the \g-ray luminosity of the
molecular clouds, this subsequently implies a significant increase in their
estimated masses\footnote{The same
level of
uncertainty would affect the $\xco$ derivation from another
total gas tracer such as the dust column-density.}. The systematic errors on
the \g-ray selection efficiency cancel out to the first
order in the estimate of the $\xco$ ratio.

The conversion factor $\xco=[1.68 \pm 0.05\,(\mathrm{stat.})\; _{-0.10}^{+0.87}
\,(\hi\;\mathrm{opacity})] \times 10^{20}$
cm$^{-2}$ (K km s$^{-1}$)$^{-1}$ is consistent with other LAT measurements in the
Local Spur, which range from 1.5 to 2 $\times 10^{20}$
cm$^{-2}$ (K km s$^{-1}$)$^{-1}$ \citep{abdo2010cascep,ackermann20113quad}. 
From the different \g-ray measurements in the Galactic plane, the $\xco$
ratio at the solar circle appears well defined. 
It is, however, significantly greater than in nearby well-resolved clouds off the plane in
Cassiopeia and Cepheus \citep{abdo2010cascep}. Whether the discrepancy is due to the sampling
resolution or to an intrinsic $\xco$ variation on different scales inside a cloud will be
investigated in the future.

Using the $\xco$ ratio, we estimated the CO-bright molecular mass in the complex. For
this purpose we considered the region at $74\degr<l<86\degr$,
$-5\degr<b<8\degr$, where most of the gas associated with the Cygnus complex is located. Assuming a
distance of $1.4$~kpc and a mean atomic weight per hydrogen
atom in the ISM of $1.36$,
we obtain a mass $2.3\,^{+1.2}_{-0.1}\times 10^{6}
M\sun$ (where the uncertainties are dominated by the $\hi$\ opacity correction). This
value (taking the different assumption on the distance into account) is
consistent with the results by \citet{schneider2006} based on higher resolution,
multi-isotopolog CO observations, and it depicts Cygnus as a super-massive molecular complex.

The small amount of CO-bright molecular gas in the outer region of the Milky Way in this
longitude window (Fig.~\ref{comaps}) makes determining its emissivities extremely
sensitive
to the details of the model (including point sources), so we do not consider it for
scientific interpretation.

\subsection{Dark neutral gas}\label{dgdisc}

\begin{figure}
\centering
\begin{tabular}{c}
\resizebox{\hsize}{!}{\includegraphics{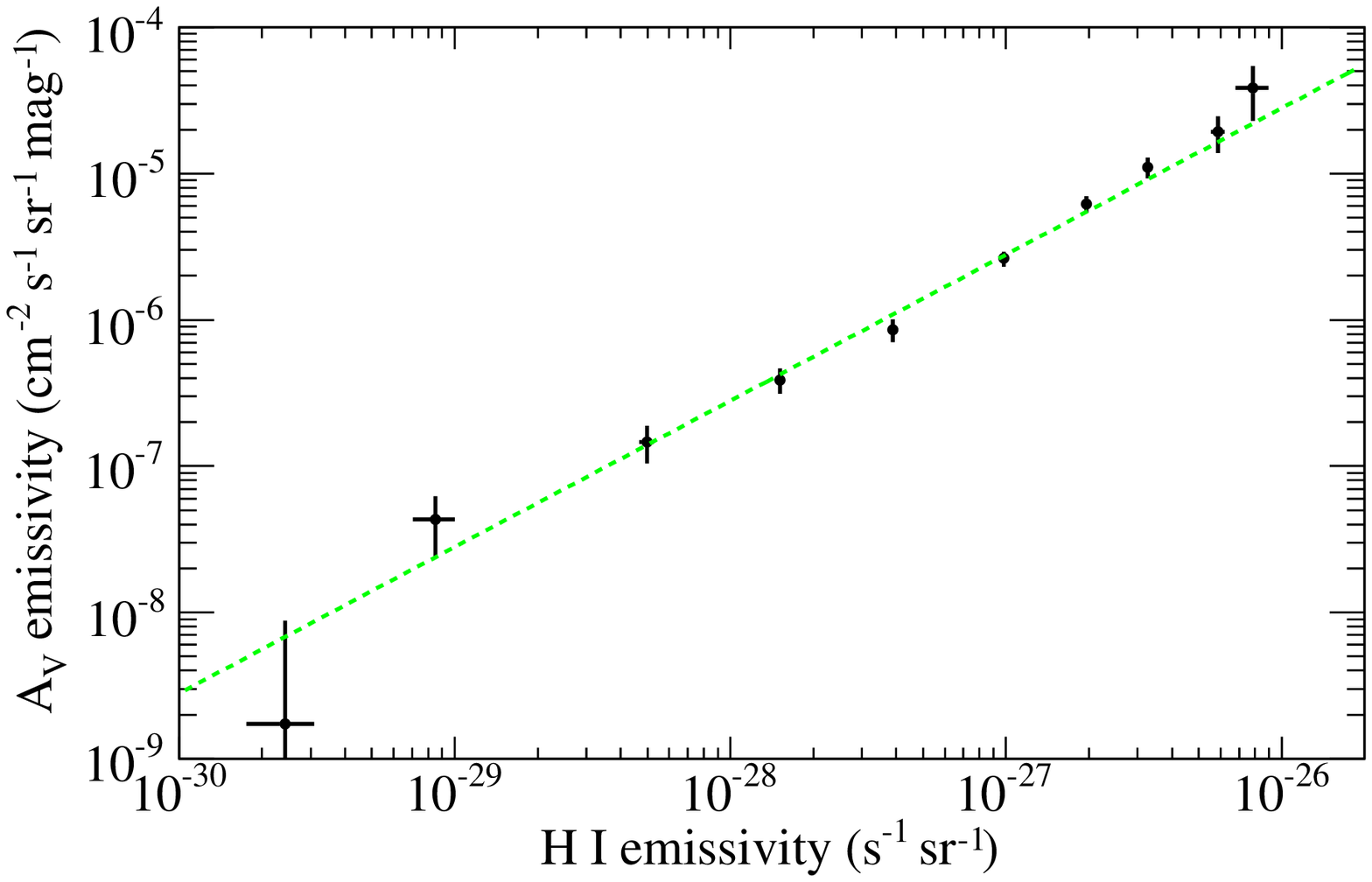}}\\
\resizebox{\hsize}{!}{\includegraphics{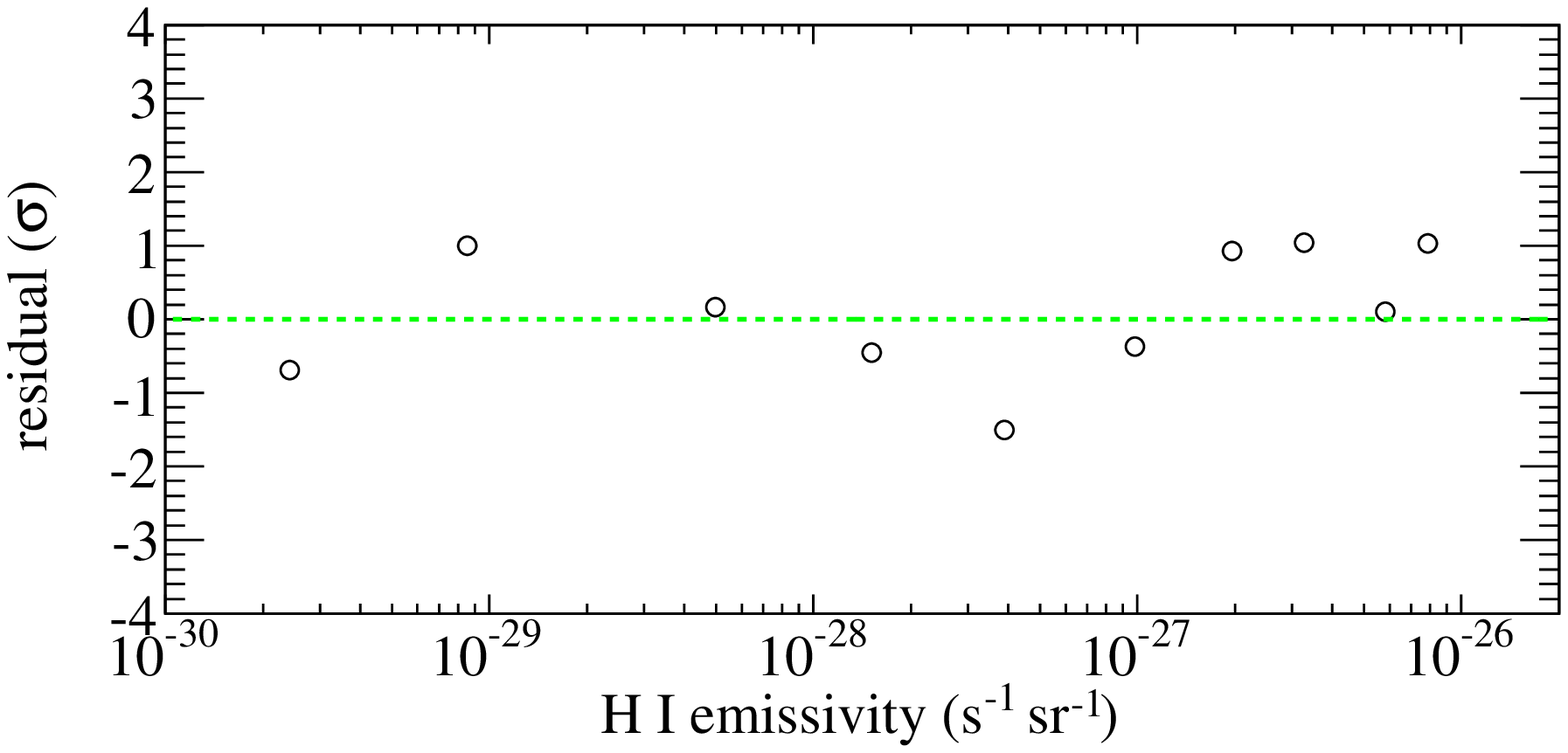}}
\end{tabular}
\caption{Top: emissivity per $\avres$ unit versus emissivity per hydrogen atom in the Cygnus
complex (for $T_S=250$~K). The points correspond to the different energy bins; the emissivities
decrease with increasing energy. 
The green line gives the best linear fit taking uncertainties on both axes into account. 
Bottom: residuals in units of standard deviations as a function of $\hi$\ emissivity.}
\label{avplots}
\end{figure}

Including the $\av$ excess map in the model corresponds to an increase of $250.6$ in the
logarithm of the likelihood (for ten additional degrees of freedom). This corresponds to a
significant
detection of \g-ray emission
associated with $\av$ excesses, formally equivalent to a $\sim 21 \sigma$ confidence level.
Figure~\ref{avplots} shows the emissivity per $\avres$ unit, $\qdust$, versus the emissivity per
hydrogen atom, $\qhinull$, in the Cygnus complex. A good linear correlation is found
between
the two emissivities over three decades in energy, proving that \g-ray emission associated with
$\av$ excesses comes from the same physical processes as that associated with $\hi$. $\av$
residuals
therefore trace interstellar gas.

With a procedure analogous to what is adopted to estimate $\xco$, we can use the emissivity per
hydrogen atom to
calibrate the dust-to-gas ratio in the dark neutral phase $\xav \equiv N(\mathrm{H})/\avres$. We
obtain $\xav=(28 \pm 2) \times 10^{20}$
cm$^{-2}$ mag$^{-1}$ in the case of $T_S=250$~K, $\xav=(48 \pm 3) \times 10^{20}$
cm$^{-2}$ mag$^{-1}$ in the case of $T_S=100$~K and $\xav=(27 \pm 2) \times 10^{20}$
cm$^{-2}$ mag$^{-1}$ in the case of optically thin medium, so $\xav$ is 
$[28 \pm
2\,(\mathrm{stat.})\; _{-1}^{+20} \,(\hi\;\mathrm{opacity})] \times 10^{20}$ cm$^{-2}$ mag$^{-1}$.

Assuming a standard total-to-selective extinction ratio $R_\mathrm{V}=\av/\ebv=3.10$
\citep{wegner2003}, the dust-to-gas ratio just estimated appears to be $\sim 50\%$ higher than the
average
value in the diffuse ISM, $N(\mathrm{H})/\ebv=58 \times 10^{20}$ cm$^{-2}$
mag$^{-1}$ \citep{bohlin1978}, and a factor of three higher than what is inferred for the
dark
phase in local clouds from \g-ray measurements, $N(\mathrm{H})/\ebv \simeq 30 \times 10^{20}$
cm$^{-2}$
mag$^{-1}$ \citep{grenier2005,abdo2010cascep}. { The discrepancy is confirmed by the
extinction data. By fitting the latter with the \hi\ and CO maps
(\S~\ref{avmaps}), we obtained
$\nhi/\av=29.6 \pm 0.1 \times
10^{20}$
cm$^{-2}$ mag$^{-1}$ (statistical error only
for $T_S=250$~K).} A possible explanation is provided by { an anomalous $R_\mathrm{V}$
ratio driven by a peculiar distribution of dust grain sizes \citep{cardelli1989}.
\citet{straizys1999} indeed} report an anomalous extinction law in the Cygnus region, showing
stronger
extinction in the violet and near UV region.

The chemical state of the dark neutral gas cannot be deduced from \g-ray observations. Whereas
there
are compelling
theoretical and observational reasons to believe that CO-quiet $\hd$ is ubiquitous in
the ISM \citep[e.g.][]{wolfire2010,magnani2003,langer2010}, we cannot exclude that part of the dark
gas
traced by $\av$ excesses is missing cold atomic gas, especially since the dark neutral phase appears
at the
interface between the atomic 
and CO-bright phases in the nearby clouds \citep{grenier2005}. 
Temperatures as low as $40-70$~K were measured in cold $\hi$\ clouds \citep{heiles2003}, and
self absorption can be strong 
when cold clouds are seen against more diffuse warm $\hi$. The $\av$
excesses in Fig.~\ref{dustmaps} partially overlap an $\hi$\ self-absorption feature
associated with the Cygnus complex \citep[][Fig.~1d]{gibson2005}.
Nevertheless, the $\hi$ to $\hd$
transition is very dynamical, both in space and time,
and it is difficult at this stage to conclude anything about the exact mix of cold dense $\hi$ and 
diffuse CO-quiet $\hd$ that forms the dark neutral phase in the outskirts of CO-bright molecular
clouds.

Regardless of its nature, the mass of the dark neutral gas in the Cygnus complex at
$1.4$~kpc amounts
to $0.9\,^{+0.4}_{-0.1}\times 10^{6}
M\sun$. { Adding} an
atomic mass of $5\,^{+4}_{-1}\times 10^{6}
M\sun$ and including the CO-bright mass estimated above the total interstellar mass of the Cygnus
complex amounts to $8\,^{+5}_{-1}\times 10^{6} M\sun$.

Assuming that all the dark neutral gas is molecular, we can calculate
the molecular dark-gas fraction $f_\mathrm{DG}=(M_\mathrm{mol}-M_\mathrm{CO})/M_\mathrm{mol}$, which
amounts to\footnote{The dark neutral gas fraction is very stable against the choice of $\hi$\
spin
temperature, therefore the error on the dark-gas fraction is only statistical.} $0.27\pm 0.02$, in
excellent
agreement
with the model by \citet{wolfire2010}. 
The dark-gas fraction is also consistent with the one by \citet{abdo2010cascep} for the nearby
Cepheus and Cassiopeia clouds, which have a factor of two lower
column
densities and masses $<2\%$ of that contained in the Cygnus complex. This also agrees
with the prediction by \citet{wolfire2010} that the dark-gas fraction is fairly
independent of the mean cloud column density and total mass for giant molecular clouds.

\section{Conclusions}

We performed an analysis of \g-ray emission across the entire Cygnus region
measured by the
\emph{Fermi} LAT in the energy range 100~MeV-100~GeV. We built a general model for the region able
to satisfactorily reproduce the LAT data. { The model includes extended sources
that have been detected over the interstellar emission model described here in association
with the Cygnus Loop and \g~Cygni supernova remnants and with a cocoon of freshly-accelerated CRs
in the innermost part of the Cygnus X
region. They are discussed in detail in companion papers.}

We measured the { average} $\xco=\nhd/\wco$ factor for clouds in Cygnus, finding a value
$[1.68 \pm
0.05\,(\mathrm{stat.})\; _{-0.10}^{+0.87}
\,(\hi\;\mathrm{opacity})] \times 10^{20}$
cm$^{-2}$ (K km s$^{-1}$)$^{-1}$ that is { well} consistent with other LAT measurements for
{ cloud
complexes} in the
Local and Perseus spiral arms \citep{abdo2010cascep,ackermann20113quad}. { These $\xco$
ratios, averaged over complexes, are, however, significantly higher than the values found at higher
sampling resolution} in
nearby clouds of the Gould Belt \citep{abdo2010cascep}.
{ Thanks to the correlation between dust and \g-ray
emission excesses, we detected the presence of conspicuous masses of dark neutral
gas not
traced by the combination of the \hi\ and CO lines, with total mass $\sim 40\%$ of the mass of the
clouds traced by CO}. The good correlation over three decades in
energy between the \g-ray emissivity per
$\av$ excess unit and per H atom strengthens the interpretation of such excesses as
{ produced by} dark
neutral
gas.
{ The
neutral gas in the Cygnus complex, combining atomic, CO-bright, and dark masses, amounts to
$8\,^{+5}_{-1}\times 10^{6} M\sun$ at a distance of 1.4 kpc.}

{ The emissivity of atomic gas  measured over the whole Cygnus complex is consistent with
other estimates in the
local interstellar space. 
We do not find evidence of any possible exclusion of CRs by
enhanced magnetic fields in the dense clouds. The
emissivity per hydrogen atom compares with LAT estimates in other regions of
the local and outer Galaxy, regardless of differences in gas surface density
by about one order of magnitude and in Galactocentric radius by $\sim 6$ kpc. This uniformity
does
not support
models based on the dynamical coupling of CRs with matter densities or predicting a strong
emissivity gradient toward the outer Galaxy.

The CR population averaged over the whole Cygnus complex { ($\sim 400$~pc)} is similar to the
Local Spur average,
in
spite of the embedded regions of conspicuous massive star-formation and potential CR accelerators.
Their impact on the CR population is only detected in the innermost region bounded by
the ionization fronts from the massive stellar clusters on a scale $< 100$~pc
\citep{cocoonScience}. No counterpart to the broadly distributed
excess of \g-ray emission seen at
energies $>10$~TeV at $65^\circ \leq l \leq
85^\circ$ \citep{abdo2007,abdo2008milagrodiff} have been detected at GeV energies so far.}

\begin{acknowledgements}
The \textit{Fermi} LAT Collaboration acknowledges generous ongoing support
from a number of agencies and institutes that have supported both the
development and the operation of the LAT, as well as scientific data analysis.
These include the National Aeronautics and Space Administration and the
Department of Energy in the United States, the Commissariat \`a l'Energie Atomique
and the Centre National de la Recherche Scientifique / Institut National de Physique
Nucl\'eaire et de Physique des Particules in France, the Agenzia Spaziale Italiana
and the Istituto Nazionale di Fisica Nucleare in Italy, the Ministry of Education,
Culture, Sports, Science and Technology (MEXT), High Energy Accelerator Research
Organization (KEK) and Japan Aerospace Exploration Agency (JAXA) in Japan, and
the K.~A.~Wallenberg Foundation, the Swedish Research Council, and the
Swedish National Space Board in Sweden.

Additional support for science analysis during the
operations phase is gratefully
acknowledged from the Istituto Nazionale di Astrofisica in Italy and the Centre National d'\'Etudes
Spatiales in France.

We made use of data from the Canadian Galactic Plane Survey (CGPS). CGPS is a Canadian project
with international partners. The Dominion Radio Astrophysical Observatory  
is operated as a national facility by the National Research Council   
of Canada. The CGPS is supported by a grant from the Natural Sciences 
and Engineering Research Council of Canada.

We thank T.~M.~Dame for providing the moment-masked CO data.

\end{acknowledgements}


\bibliographystyle{aa}

\end{document}